\documentclass[a4paper,11pt]{article}
\pdfoutput=1 

\usepackage{jinstpub} 

\usepackage[utf8]{inputenc}

\usepackage{lineno}
\usepackage{color}

\usepackage[outdir=./]{epstopdf}

\title{\boldmath Commissioning of the highly granular SiW-ECAL technological prototype}

\author{\centering S.\,Bilokin\footnote{ Now at IPHC, Strasbourg.}, J.\,Bonis, P.\,Cornebise, A.\,Gallas, A.\,Irles**,  R.\,P\"oschl, F.\,Richard, A.\,Thiebault, D.
\,Zerwas \\ {\it Laboratoire de l'Acc\'elerateur Lin\'eaire, CNRS/IN2P3 et Universit\'e de Paris-Sud XI, Centre Scientifique d'Orsay B\^atiment 200, BP 34, F-91898 Orsay 
CEDEX, France } 
\\ M.\,Anduze, V.\,Balagura, V.\,Boudry, J-C.\,Brient, E.\,Edy, G.\,Fayolle, M.\,Frotin, F.\,Gastaldi, A.\,Lobanov, F.\,Magniette, J.\,Nanni, M.\,Rubio-
Roy \footnote{ Now at Spintec, Grenoble} , K.\,Shpak, H.\,Videau, D.\,Yu\footnote{ Now at IHEP, Beijing.} \\ {\it Laboratoire Leprince-Ringuet (LLR) -- \'{E}cole Polytechnique, 
CNRS/IN2P3, Palaiseau, F-91128 France } 
\\ S.\,Callier, F.\,Dulucq, Ch.\,de la Taille, N.\,Seguin-Moreau \\ {\it Laboratoire OMEGA -- \'{E}cole Polytechnique-CNRS/
IN2P3, Palaiseau, F-91128 France } 
\\ J.E.\,Augustin, R.\,Cornat, J.\,David, P.\,Ghislain, D.\,Lacour, L.\,Lavergne\footnote{ Now at IRAP, Toulouse.}, J.M.\, Parraud \\ {\it Laboratoire de Physique Nucl\'eaire et de Hautes Energies 
(LPNHE), Sorbonne Universit\'e, UPD, CNRS/IN2P3, 4 Place Jussieu, 75005 Paris, France } 
\\ K.\,Kawagoe, Y.\,Miura, I.\,Sekiya, T.\,Suehara, T.\,Yoshioka, \\ {\it Department of Physics and Research Center for Advanced Particle Physics, 
 Kyushu University, 744 Motooka, Nishi-ku, Fukuoka 819-0395, Japan } 
\\ D.\,Jeans \\ {\it Institute of Particle and Nuclear Studies, KEK, 1-1 Oho, Tsukuba, Ibaraki 305-0801, Japan } 
\\ J.\,S.\,Chai \\ {\it Department of Electrical and Computer Engineering, Sungkyunkwan Universtity, 16419, Suwon, Gyeonggi-do, Korea}}

\emailAdd{irles@lal.in2p3.fr}

\abstract{
In this article we describe the commissioning and a first analysis of the the beam test performance of a small prototype of a highly granular silicon tungsten calorimeter. The prototype features detector elements with a channel number similar to that envisaged for e.g. the ILD Detector of the International Linear Collider (ILC). The analysis demonstrates the capability of the detector to record signals as low as 0.5 MIP. Further, no loss of performance has been observed when operating the detector in a high magnetic field.   
}


\begin{document}
\maketitle
\flushbottom

\section{Introduction}

Future accelerator based particle physics experiments
require very precise and detailed reconstruction of the final states produced
in the beam collisions. A particular example is the next generation of $e^{+}e^{-}$
linear colliders such the ILC\cite{Behnke:2013xla,Baer:2013cma,Adolphsen:2013jya,Adolphsen:2013kya,Behnke:2013lya}.
This project will provide collisions of polarized beams with center-of-mass energies of 250 GeV to 1 TeV
studied by multipurpose detectors. Two projects are proposed:
the International Large Detector (ILD) and the Silicon Detector (SiD)\cite{Behnke:2013lya}.
To meet the precision required by the ILC 
physics goals, event reconstruction is based on the 
Particle Flow (PF) technique \cite{Brient:2002gh,Morgunov:2004ed,Sefkow:2015hna}.
PF implies the reconstruction of every single particle of the final state. 
This in turn requires the construction of calorimeters with unprecedented granularity.

The CALICE collaboration is driving most of the efforts on R\&D of highly granular calorimeters
for future linear colliders by investigating and building prototypes for several
calorimeter concepts. One of these calorimeters 
is the silicon-tungsten electromagnetic calorimeter, SiW-ECAL.
Sensors made of pixellated Silicon diodes (Si) are sandwiched between tungsten (W) absorbers.
The combination of Si and W choices  makes possible the design and construction
of a very compact calorimeter with highly granular and compact active layers.
In the detectors at the ILC, the very-front-end (VFE) electronics of the calorimeters will be
embedded in the detector units. 
The desired signal dynamic range in each channel goes from 0.5 MIP to 3000 MIPs, where the MIP acronym stands 
for the energy deposited by a minimum-iononizing-particle.
To reduce overall power consumption, the SiW-ECAL will exploit the special bunch structure
foreseen for the ILC with the $e^{+}e^{-}$ bunches trains grouped in spills
of $\sim$ 1-2 ms width separated by $\sim$ 200 ms. 
The front-end electronics will only be enabled during the spills and the bias currents will be shut down between the spills. 
This technique is usually denominated power pulsing. 
The calorimeters are operated in self-trigger mode (each channel featuring an internal trigger decision chain) and zero suppression mode. 

\section{The SiW-ECAL technological prototype}

The first SiW-ECAL prototype was the so called SiW-ECAL physics prototype.
It was extensively tested between 2005 and 2011 at DESY, FNAL and CERN ~\cite{Adloff:2011ha,Anduze:2008hq,Adloff:2008aa,Adloff:2010xj,CALICE:2011aa,Bilki:2014uep}. 
For the physics prototype, the VFE was placed outside the active area with no particular constraints in the power consumption.
It consisted of 30 layers of Si as active material alternated with tungsten plates as absorber material.
The active layers were made of a matrix of 3x3 Si wafers of 500 $\mu$m thickness. Each of these wafers was segmented in matrices of
6x6 squared channels of 10x10 mm$^{2}$.
The prototype was divided in 3 modules of 10 layers with different W depth per layer in each of these modules
(0.4, 1.6 and 2.4 $X_{0}$) making a total of 24 $X_{0}$.
That very first prototype offered a signal over noise on the measured charge of 7.5 for MIP like 
particles.

The current prototype is called the SiW-ECAL technological prototype. It addresses the main technological challenges: compactness,
power consumption reduction through power pulsing and VFE inside the detector.
The study presented here extends considerably the work presented in Ref. \cite{Amjad:2014tha} 
on an earlier version of the technological prototype that comprised 
e.g. four times less channels per layer. 
Further, in this earlier study the detector was still operated in continuous power mode.
In this section we described in detail
the main features and characteristics of the technological prototype.

\subsection{Silicon sensors}
\label{sec:wafers}

The sensors consist of high resistivity (bigger than 5000 $\Omega\cdot$cm)
silicon wafers with a thickness of $320\pm15\mu$m.
The size of the wafers is $9\times9$ cm$^{2}$ and each of them is subdivided in an array of 256 PIN diodes of $5\times5$ mm$^{2}$.
A MIP traversing the PIN parallel to its normal will create $\sim$ 80 $h^{+}e^{-}$ pairs per $\mu$m which corresponds to 4.1 fC.
The original design of the silicon wafers included an edge termination made of floating guard-rings.
It was observed in beam tests \cite{Cornat:2015eoa,Cornat:2009zz} that the capacitive coupling between such floating guard-rings 
and the channels at the edge created not negligible rates of parasitic signals.
events in tests with high energy beams (pions and electrons with energies larger than 20-40 GeV)
An R\&D program together with Hamamatsu Photonics (HPK Japan) was conducted to study the guard-ring design 
as well as the internal crosstalk. It was concluded that using wafers without guard-ring and with a width of the peripheral areas lower than 
500$\mu$m thanks to the use of stealth dicing technique, the amount of these squared events 
can be reduced to be at negligible level.
For the setup described in this article we used different solutions for the edge terminations.
For all of them, the expected rate of fake events is however negligible due to
the low energy of the particle beam studied in the present article.

\subsection{SKIROC: Silicon pin Kalorimeter Integrated ReadOut Chip}
\label{sec:skiroc}

\begin{figure}[!ht]
  \centering
    \includegraphics[width=4in]{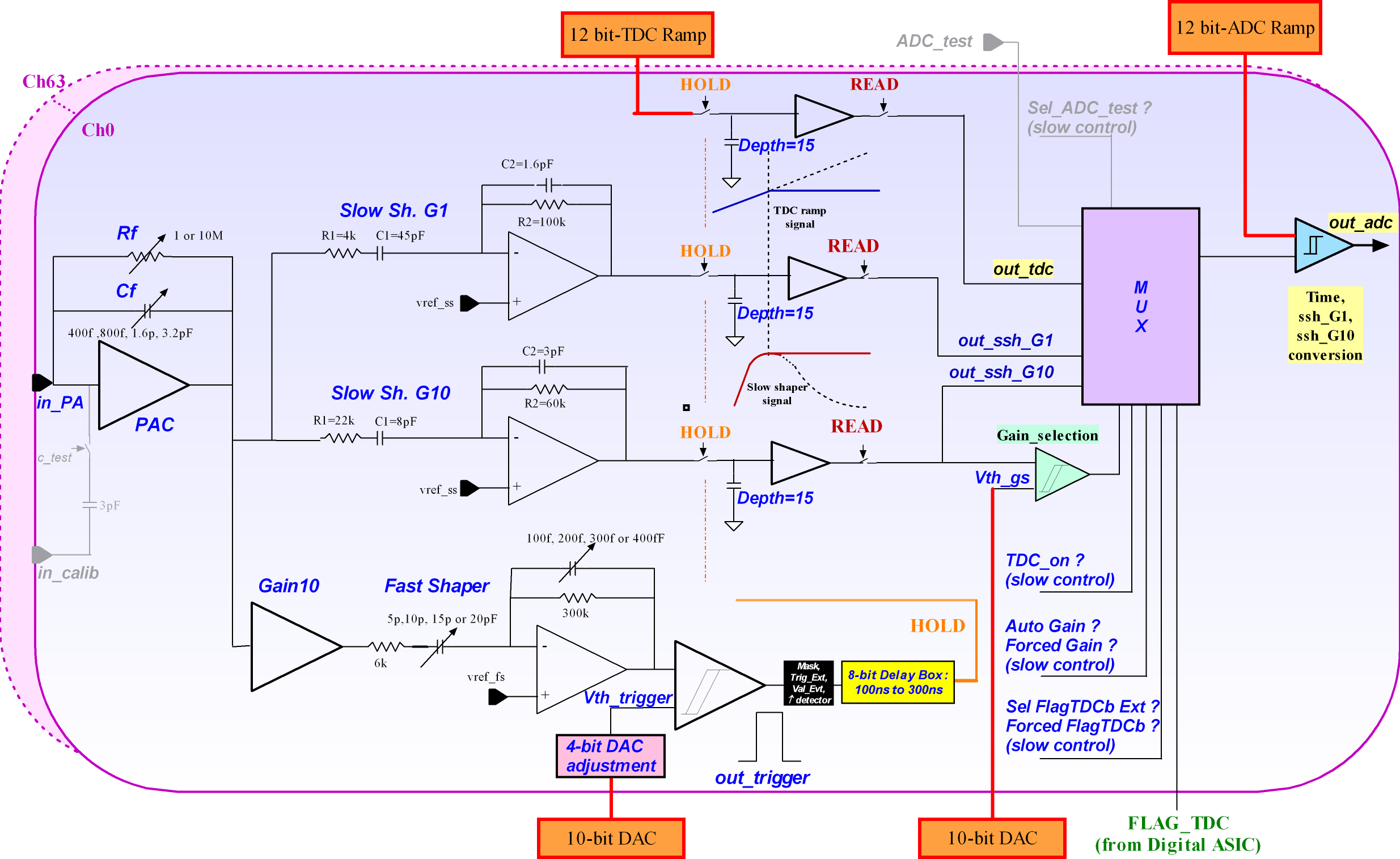}
\caption{The schematics of the analog part of SKIROC2.}
\label{SKIROC2}
\end{figure}

The SKIROC\cite{Callier:2011zz} (Silicon pin Kalorimeter Integrated ReadOut Chip) is a
very front end ASIC (application-specific integrated circuits)
designed for the readout of silicon PIN diodes.
In its version SKIROC2 it consists of 64 channels in AMS 0.35 $\mu$m SiGe technology.
A schematic view of the analog part of the SKIROC2 is shown in Figure \ref{SKIROC2}.
Each channel comprises a low noise charge preamplifier of variable gain followed by two branches:
a fast shaper for the trigger decision and a set of dual gain slow shapers for charge measurement.
The gains can be controlled by modifying the feedback capacitance.
When there is a trigger, the charge of the triggered channel is held after a fixed delay 
and stored in one of the memory cells of the 15 cell deep physical switched capacitor array (SCA). 
At the end of the slow clock period in which a trigger occurred
the held of the charge of all the channels not triggered is launched.
Finally, the signals are digitized by a Wilkinson type analogue to digital converter.
The digitized information includes tags to identify the channels with and without trigger.

The SKIROC ASICs can be power-pulsed. 
Under conditions as designed for the ILC the power consumption is expected to be as low as 25muW.
The power pulsing feature is used for all the results discussed in this paper.

\subsection{Active Sensor Units}
\label{sec:ASU}

\begin{figure}[!t]
  \centering
  \begin{tabular}{l}
    \includegraphics[width=4in]{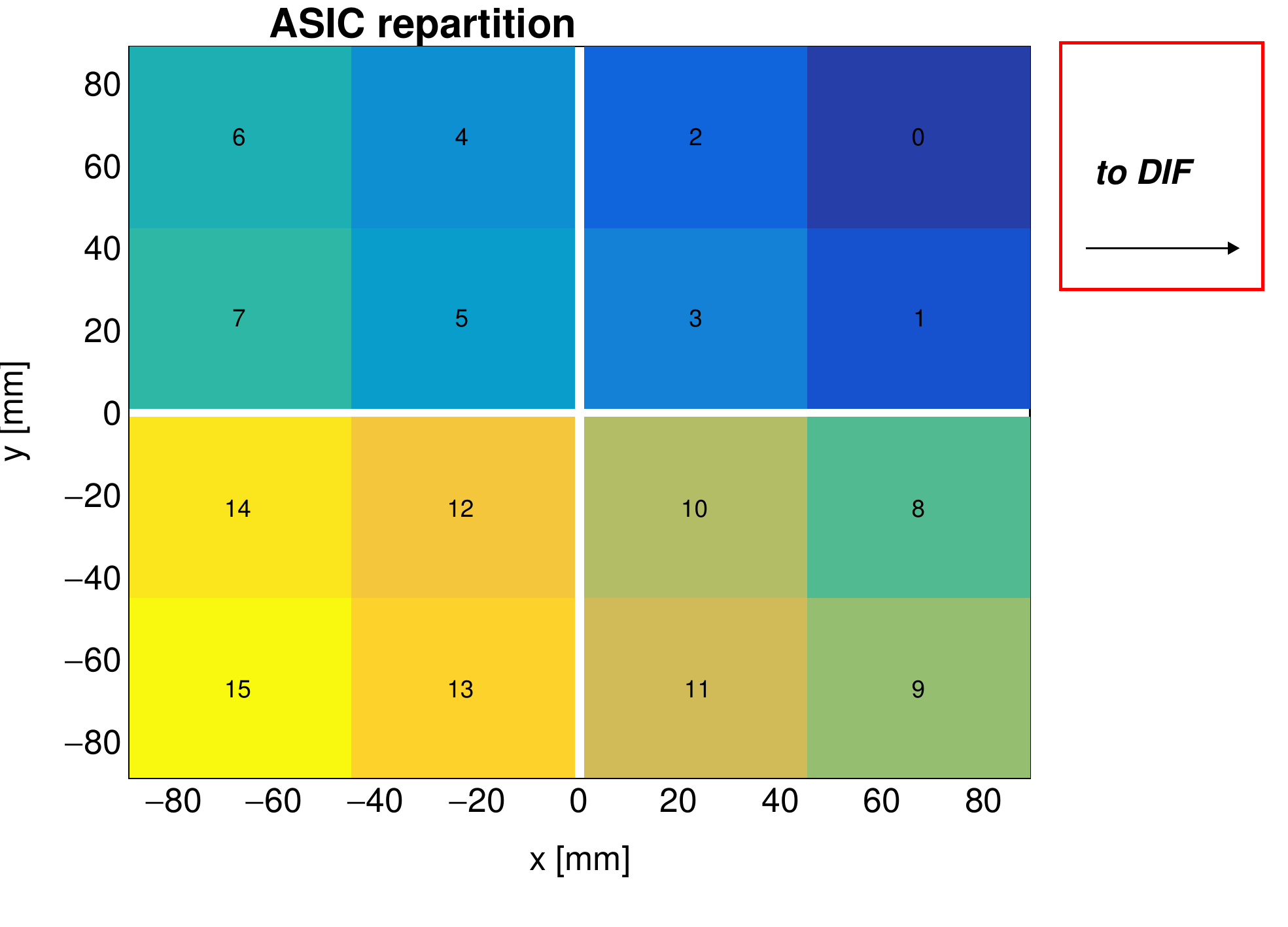}  \\
    \includegraphics[width=4in]{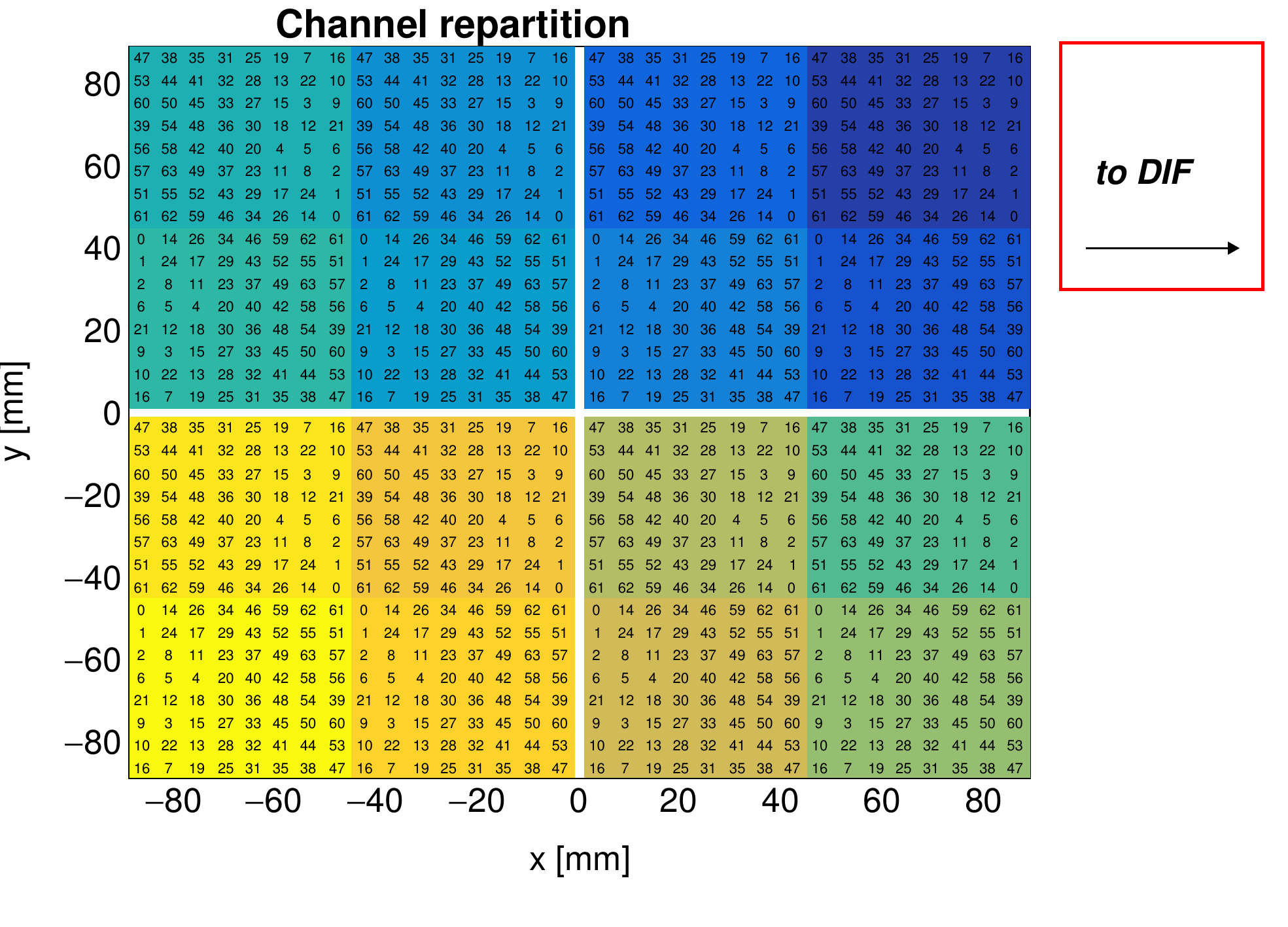}  \\
  \end{tabular}
  \caption{Repartition of the ASIC (up) and channels (down) in one ASU as viewed from the PCB side.
    The channels are separated (in x and y) by 5.5 mm$^{2}$.
    The empty cross in the middle of the ASU corresponds to the 1 mm separation between the sensors.
    The areas covered by the different ASICs and channels
    are labeled with numbers following design and DAQ criteria: from 0-16 in the case of the ASICs and from 0-63 in the case of the channels.
  }
\label{ASU}
\end{figure}

The entity
of sensors, thin PCB (printed circuit boards) and ASICs is called Active Signal Units or ASU.
An individual ASU has a lateral dimension of 18x18 cm$^{2}$.
The ASUs are currently equipped
further with 16 SKIROC2 ASICs for the read out and feature 1024 square pads (64 per ASIC) of 5.5x5.5 mm.
The channels and ASICs are distributed along the ASU as shown in Figure \ref{ASU}. Each ASU is equipped with 4 silicon wafers as the described in Section \ref{sec:wafers}.
The high voltage is delivered to the wafers using a HV-kapton sheet that covers the full extension of the wafers.

The current version of the PCB is called the FEV11. It has a thickness of 1.6 mm which
grows up to 2.7 mm when the ASICs in its current packaging (1.1 mm thick LFBGA package) are
bonded in top of it.

\subsection{Data AcQuisition system}
\label{sec:DAQ}

The subsequent chain of the data acquisition (DAQ)\cite{Gastaldi:2014vaa} system consists of three components.
They are enumerated from upstream to downstream from the data flow perspective:

\begin{enumerate}
\item The first component is the so called detector interface (DIF) which is placed at the beginning of each layer.
\item Each DIF is connected via a standard HDMI to the  second component: a concentrator card called the Gigabit Concentrator Cards (GDCCs). These cards control up to 7 DIFs. They collect all data from the DIFs and distribute in turn the system clock and fast commands.
  \item The most downstream component, is the clock and control card (CCC) which
provides a clock, the control fan-out of up to 8 GDCCs and accepts and distributes external signals (i.e. signals
generated external pulse generator to simulate the ILC spill conditions).
\end{enumerate}

The whole system is controlled by the Calicoes and the Pyrame DAQ software version 3~\cite{Rubio-Roy:2017ere,Magniette:2018wdz}.

\subsection{The prototype setup}
\label{sec:setup}

\begin{figure}[!ht]
  \centering
    \includegraphics[width=6in]{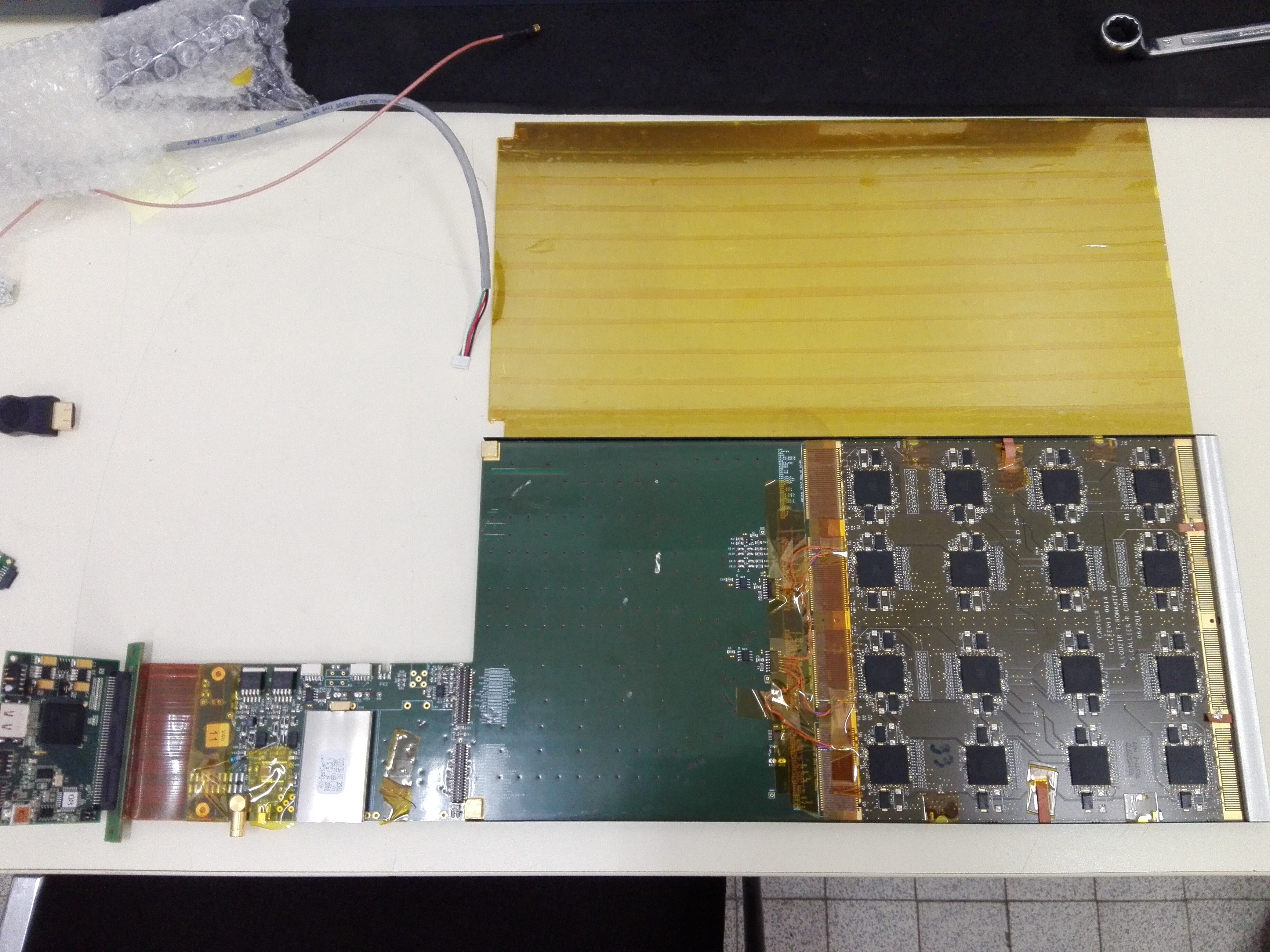} 
  \caption{Open single readout layer with FEV11 ASU, 16 SKIROC 2 and the interface card visibles. }
\label{ASU2}
\end{figure}

A fully equipped readout module is shown in Fig. \ref{ASU2}. These modules consist in this case of one ASU
connected to a data acquisition system (DAQ) through an adapter board, called SMBv4.
The ASU is the squared PCB on the right side, equipped with the 16 SKIROC2 bump bonded.
It is connected with kapton connectors to the SMBv4 PCB which occupies the rest of the picture.
The SMBv4 also holds other
services e.g. the super capacitance used for the power pulsing. 
This decoupling capacitor (seen in the left bottom part of the photograph)
 of 400mF with 16 m$\Omega$ of equivalent serial resistance
provides enough local storage 
of power to assure stable low voltage supply during the power pulsing. 
The readout modules are embedded on a "U" shaped carbon structure to protect the wafers.
The full system is then covered by two aluminum plates
to provide electromagnetic shielding and mechanical stability.

\begin{figure}[!t]
\centering
\begin{tabular}{l}
\includegraphics[width=6.0in]{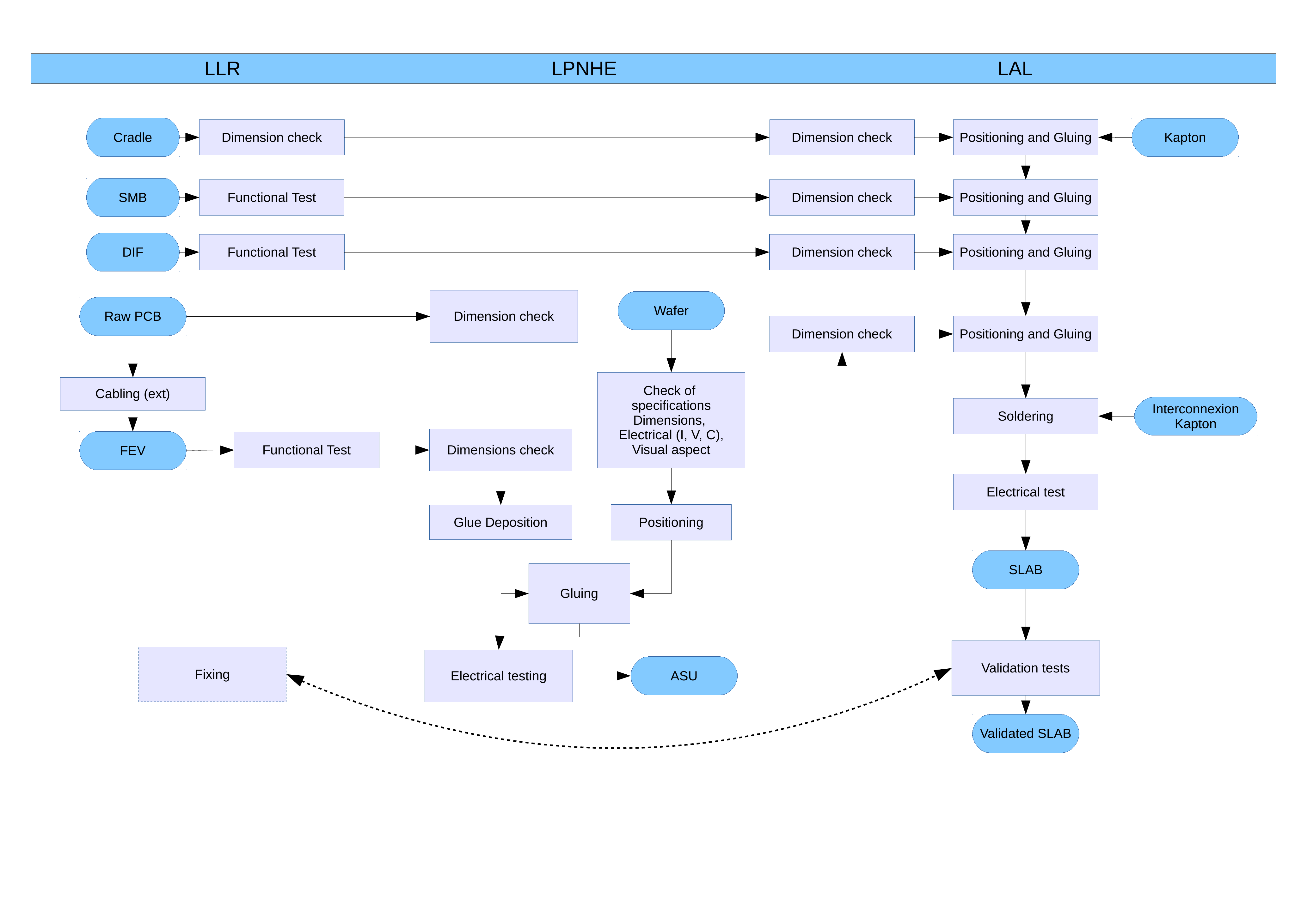}
\end{tabular}
\caption{Process flow for the assembly of the SiW-ECAL readout layers.}
\label{assembly}
\end{figure}

For the production of the small sample of readout modules studied in this document,
a scalable working procedure has been established among several groups \cite{Boudry:2318814}
profiting from the funding of projects like AIDA2020 or the HIGHTEC emblematic project
of the P2IO. A schematic view of this assembly procedure chain can be seen in
Figure \ref{assembly}. For more details we refer to Ref.\cite{Boudry:2318814}.

\begin{figure}[!ht]
\centering
\begin{tabular}{l}
\includegraphics[width=4.0in]{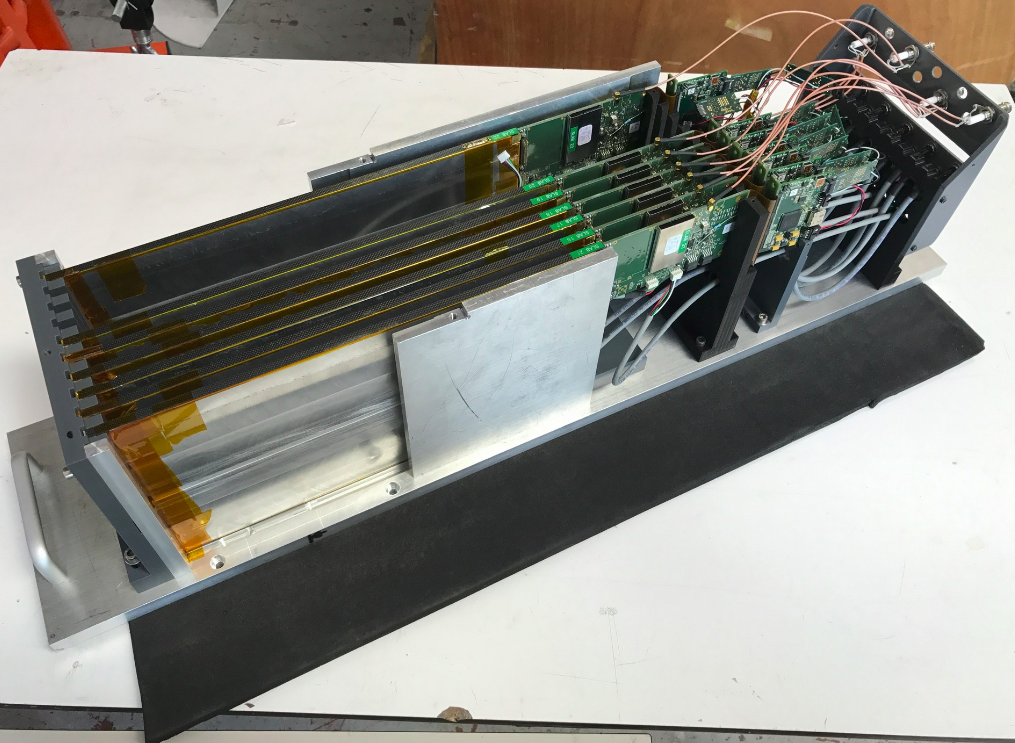} 
\end{tabular}
\caption{Prototype with 7 layers inside the aluminum stack.}
\label{proto}
\end{figure}

A picture showing the SiW-ECAL technological prototype setup can be seen in Figure \ref{proto}.
The current prototype consists of 7 layers of 
readout layers housed in a PVC and aluminum structure that can host up to 10 layers in slots separated by 15 mm each.
The first six layers were placed in the first six slots and the last one was in the last slot.
In the following sections, we will refer to layers number 1 to 7, where
the 1 is the most upstream of the beam.
This setup is used for commissioning (Section \ref{sec:commissioning}) and for the beam test
(Section \ref{sec:beamtest}).

\section{Commissioning}
\label{sec:commissioning}

Earlier experiences with the SKIROC2 ASIC are reported in
Refs. \cite{Amjad:2014tha,Suehara:2018mqk}). 
Internal SKIROC2 parameters reported in these references are adopted in the following
unless stated otherwise.
For example, a gain value of 1.2pF for the preamplifier is used. 
With this gain, the SKIROC2 features a linearity better than 90\% 
for 0.5-200 MIPs, which is sufficient for 
electromagnetic showers created by few GeV 
electrons or positrons.

The main goal of the the commissioning  procedure is 
the optimization of the trigger thresholds to levels in which we are able
to record physics signals bellow the MIP level without saturating our DAQ with
noise signals. This requires a careful and systematic procedure to:

\begin{enumerate}
  \item identify the readout channels that are noisy in high trigger threshold above MIP signal conditions;
  \item and select the optimal trigger threshold levels.
\end{enumerate}

During the commissioning, we observed the repetition of coherent noise events affecting to several
readout layers at the end of acquisitions with long gating time. The situation
could be remedied by improving the isolation of the individual readout layers and by reducing the
data taking to short gating times.
In any case, all runs dedicated to the commissioning are usually characterized by their short gating windows for the acquisition (1-2ms)
at low repetition frequencies (1-5 Hz)
to minimize the chances of having real events due to cosmic
rays during the data taking.

\subsection{Tagging and control of the noisy channels.}
\label{sec:comm_noise}

The list of the noisy channels was
obtained by means of dedicated data taking runs. In these runs
we scan relatively high trigger thresholds from larger to lower values ({\it i.e.} between 1-3 MIPs) and progressively mask channels that exhibit counts. 
In each step, the decision of tagging a channel as noisy was taking following the rules described next: 

\begin{itemize}
\item if the channel was triggered at rates larger than 0.5-1\% of the total number of triggers per ASIC it was added to the list;
\item if a channel was tagged as noisy in, at least, three of the readout layers, it was tagged as noisy for all and added to the list of channels being suspect of suffer from routing issues.
\end{itemize}

Following this procedure, we found two different types of noisy channels. One set consists of
channels randomly distributed along the surface of every ASU and the other consists
of channels located in specific areas and systematically noisy in all the ASUs. Preliminary inspections of the PCB layout
hint that the channels in the latter set may be noisy due to
improvable routing of the PCB.
Deeper studies on the PCB routing must be conducted to clarify this.
All the noisy channels have been identified and masked and the power
of their preamplifiers has been disabled. All the results shown in the following
sections are obtained in these conditions.

\begin{figure}[!t]
  \centering
  \includegraphics[width=4in]{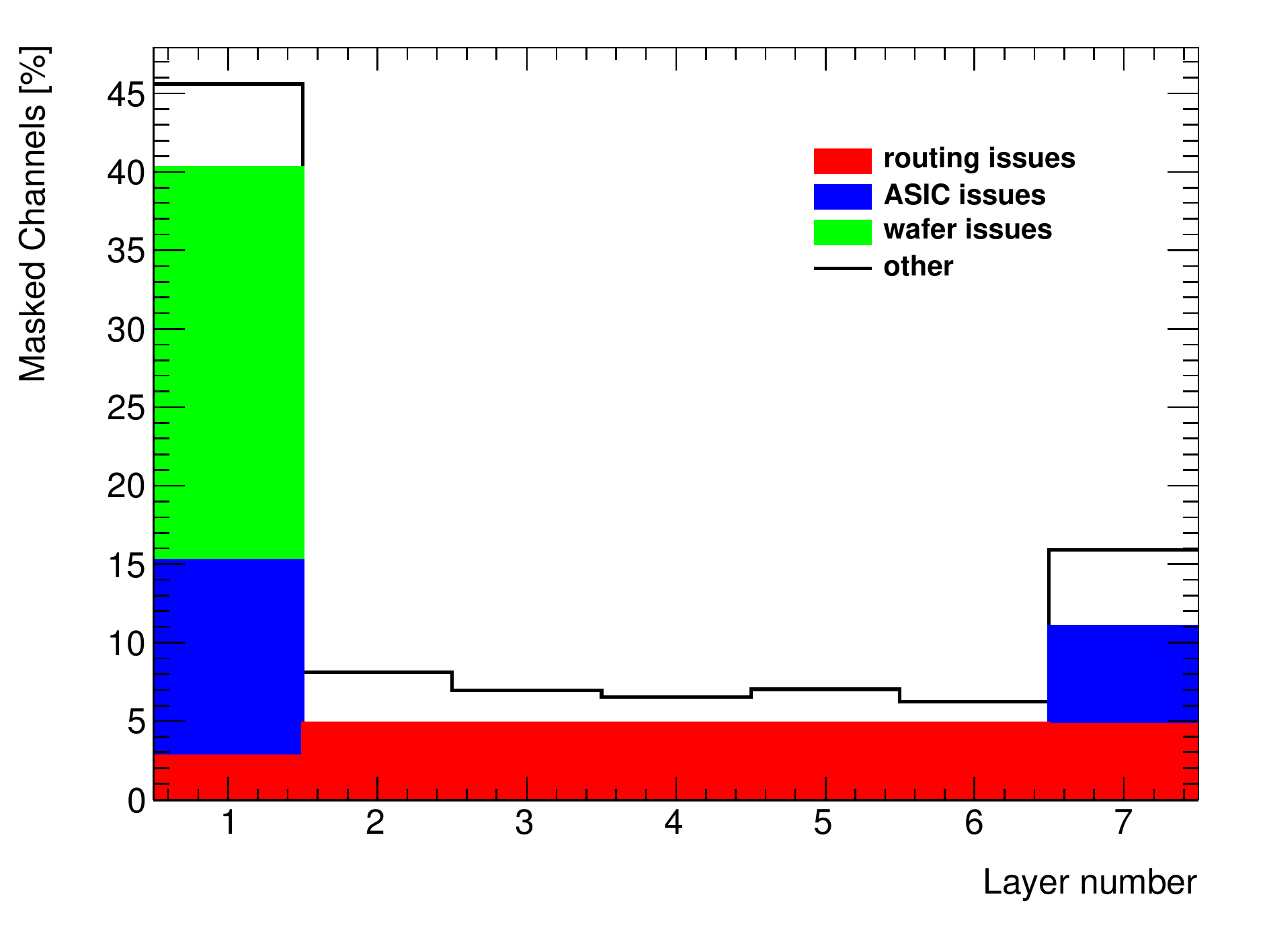} \\
\caption{Fraction of channels that are tagged as noisy in all readout layers.}
\label{noisycells}
\end{figure}

In addition to the different noisy channel types described above, we also have
masked full sectors of the readout layers if an ASIC was tagged as faulty (at least 70\% of channels 
listed as noisy) or if a Si-wafer was damaged (high leakage currents).
The results of this study are summarized in Figure \ref{noisycells}.

\subsection{Optimal trigger threshold determination}
\label{sec:comm_trigger}

\begin{figure}[!ht]
  \centering
  \begin{tabular}{ll}
  \includegraphics[width=2.8in]{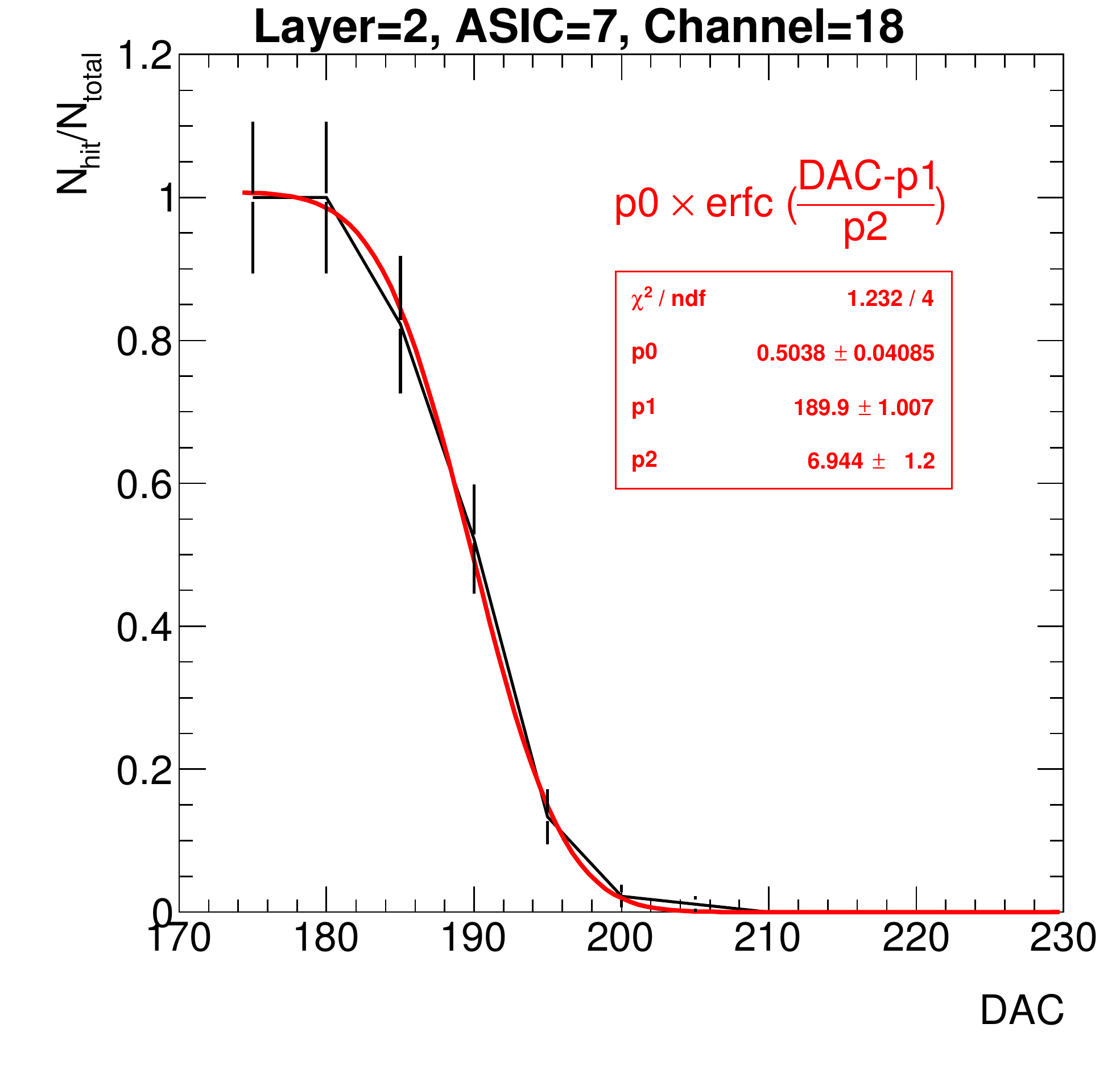} & \includegraphics[width=2.8in]{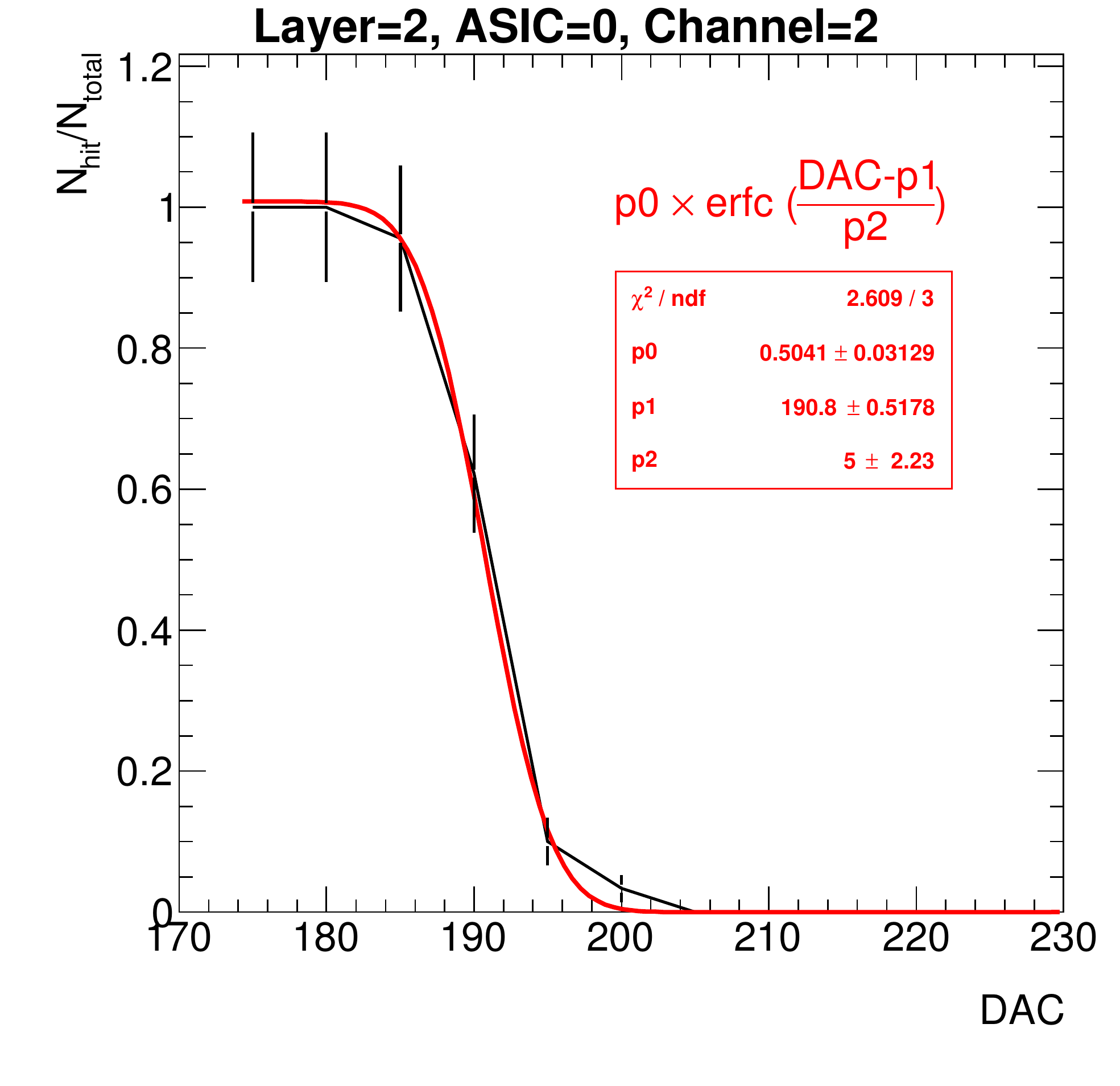}
  \end{tabular}
\caption{Two threshold scan curves.}
\label{scurve_channels}
\end{figure}

After the noisy channels have been masked, dedicated trigger threshold
scans are performed. The results are presented with the so-called the threshold scan curves where the
x-axis represents the threshold value and the y-axis
the number of recorded signals normalized to 1. The threshold values are given in internal DAC units
which are translated to meaningful physical quantities in Section \ref{sec:comm_trigger_sn}.
In the absence of external signals (cosmic rays, injected signals, etc) 
the falling edge position in the threshold scan curves
is due to the electronic noise
at the output of the fast shaper (the trigger decision branch on the SKIROC)
and it depends on the slow clock frequency.
These threshold scan curves are approximated by a complementary error function:

\begin{equation}
\frac{2p_{0}}{\sqrt(\pi)} \int_{\frac{DAC-p_{1}}{p_{2}}}^{\infty} e^{-t^{2}} dt,
\label{eq_S-curve}
\end{equation}
where $p_{0}$ is 1/2 of the normalization, $p_{1}$ is the value in which the noise levels are 
the 50\% of its maximum and $p_{2}$ give us the width of the error function. 
In Figure \ref{scurve_channels} 
two threshold scans curves are shown together with the fit by
the theoretical function.

For every ASICs, after performing the fit of the theoretical curves
to the threshold scans, the average values of the   $p_{1}$ and $p_{2}$ are calculated.
These are represented by $<p_{1,2}^{ASIC}>$ in the following.
The optimal threshold value of every ASIC, in DAC units, was chosen 
using the following formula
\begin{equation}
DAC_{optimal}^{ASIC} =maximum(<p_{1}^{ASIC}> + 5 \times <p_{2}^{ASIC}>,230).
\end{equation}
This formula was applied 
if at least the the 30\% of the 64 channels in the ASIC could be fitted.
If not, a global threshold value of 250 was set.

The optimal trigger threshold values for all ASICs are shown in Figure \ref{trigger_thresholds}, in internal DAC units and in MIPs. In the next section we explain how the conversion is done.

\subsection{S/N ratio for the trigger decission}
\label{sec:comm_trigger_sn}

\begin{figure}[!t]
    \centering
  \begin{tabular}{l}
    \includegraphics[width=4in]{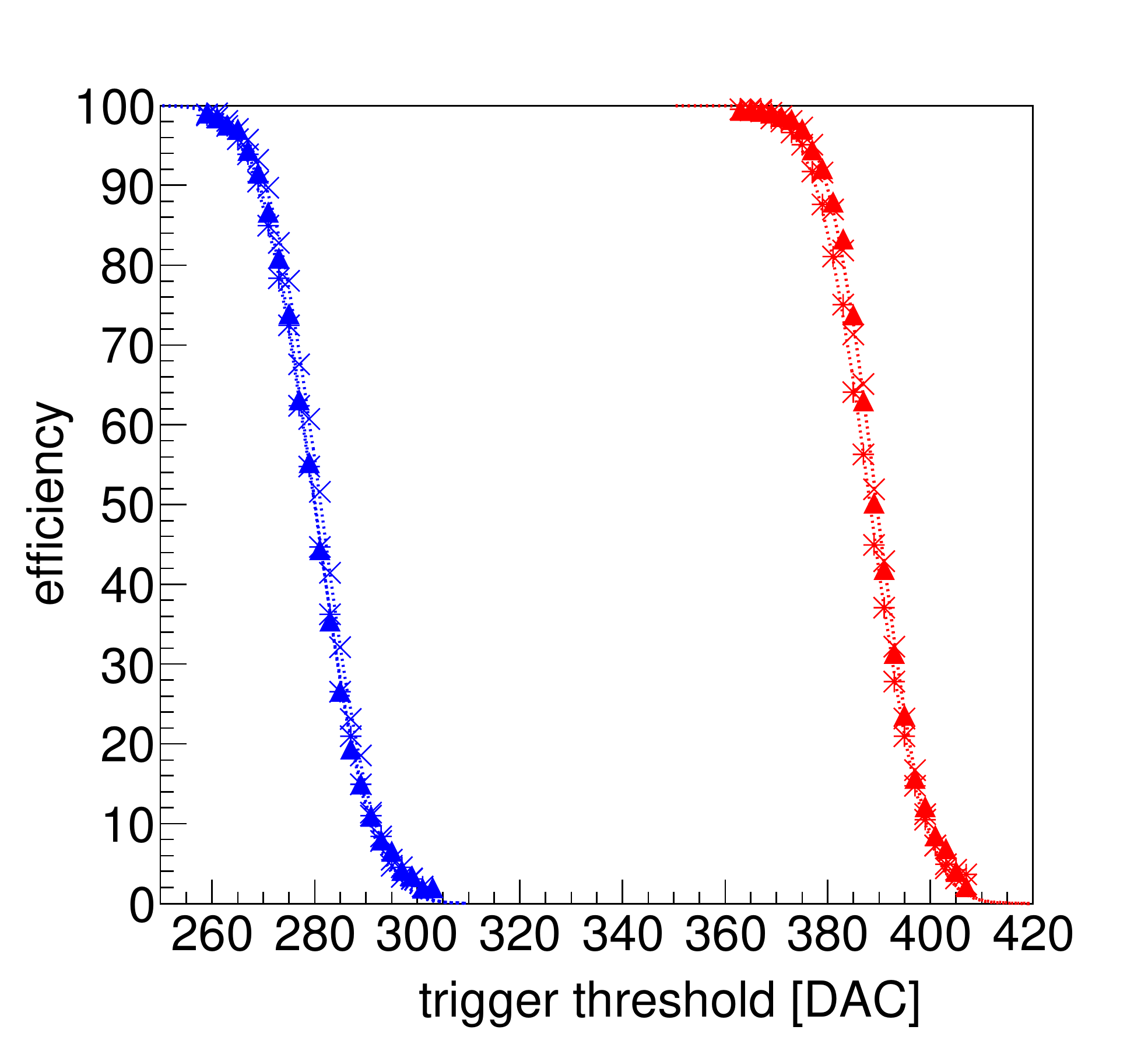} \\
	\end{tabular}
\caption{Threshold scan curves with charge injection (1 MIP in blue and 2 MIPs in red) for two different channels in a SKIROC2 testboard.}
\label{scurves_injection}
\end{figure}

Performing threshold scans using
real signals allows to calculate the signal over noise (S/N) ratio
for the trigger decision.
For that we compare the curves for 1 MIP and 2 MIP injected signals.
The S/N is, in the following, defined as the ratio between the distance of both curves at its
50\% and the width of the curves. The width is defined as the half of distance, in threshold units, between the
curve at $50\pm34\%$.
In Figure \ref{scurves_injection} we see the 1 MIP and 2 MIP curves obtained for several channel
in a SKIROC testboard in which a single SKIROC2 in BGA package is placed and the 1 MIP and 2 MIPs 
size signals are directly injected in the preamplifier 
(via a 3 pF capacitor located in the injection line as shown in Figure \ref{SKIROC2}).

\begin{figure}[!ht]
    \centering
  \begin{tabular}{l}
	\includegraphics[width=4in]{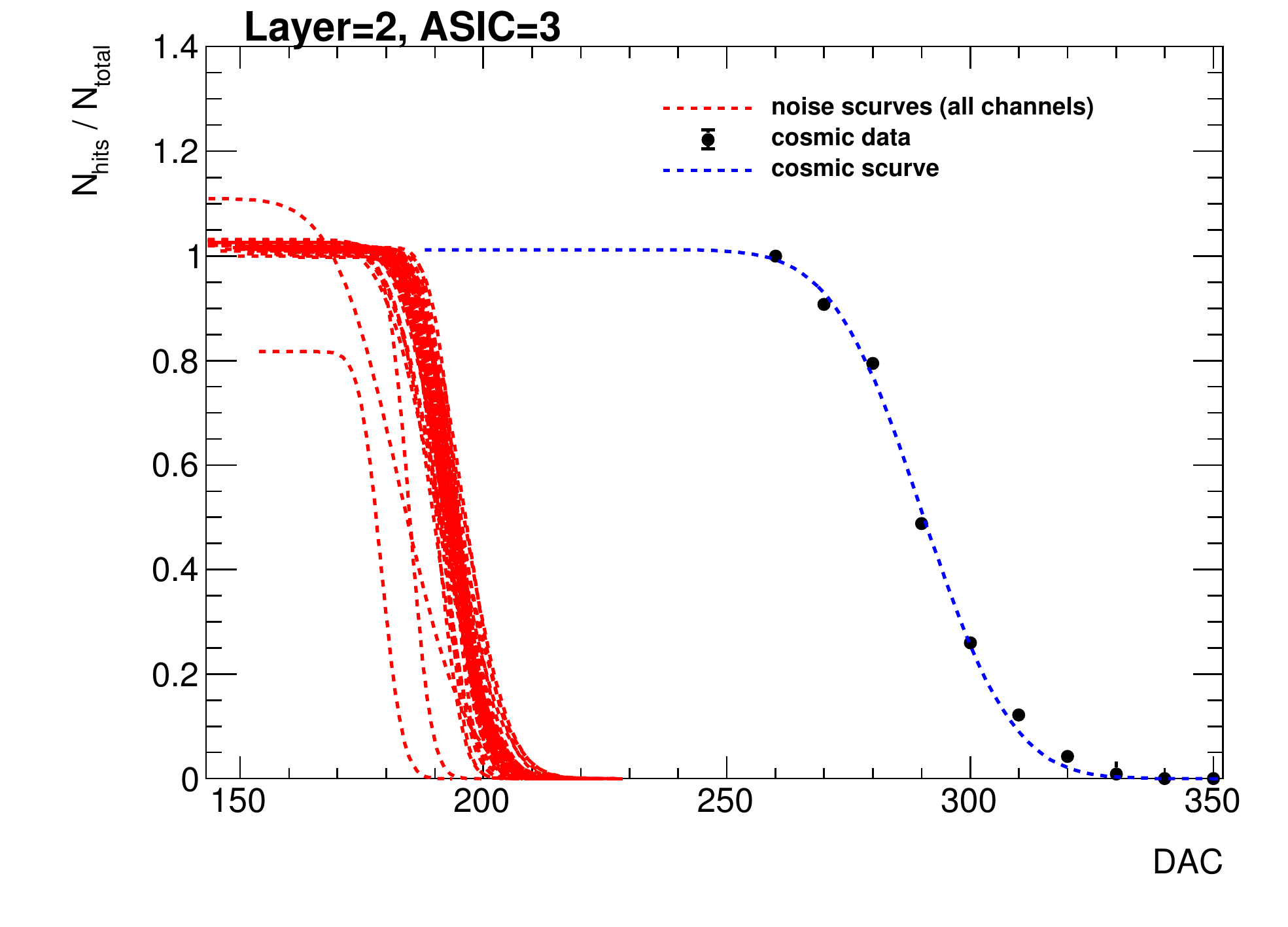} 
	\end{tabular}
\caption{Threshold scan curves for noise (channel by channel, only the result of the fit) and cosmic rays (all channels together) for one ASIC in layer 2.}
\label{scurves_cosmics}
\end{figure}

We have obtained similar results using real signals, in this case cosmic rays signals. This is shown 
in Figure \ref{scurves_cosmics}  where 
we show the result of the fit to the threshold scan curves cosmic rays integrated for all channels
in one ASIC. For completeness, the fit of threshold scan curves for all channels
in the same ASIC are also shown.

\begin{figure}[!t]
  \centering
  \includegraphics[width=4in]{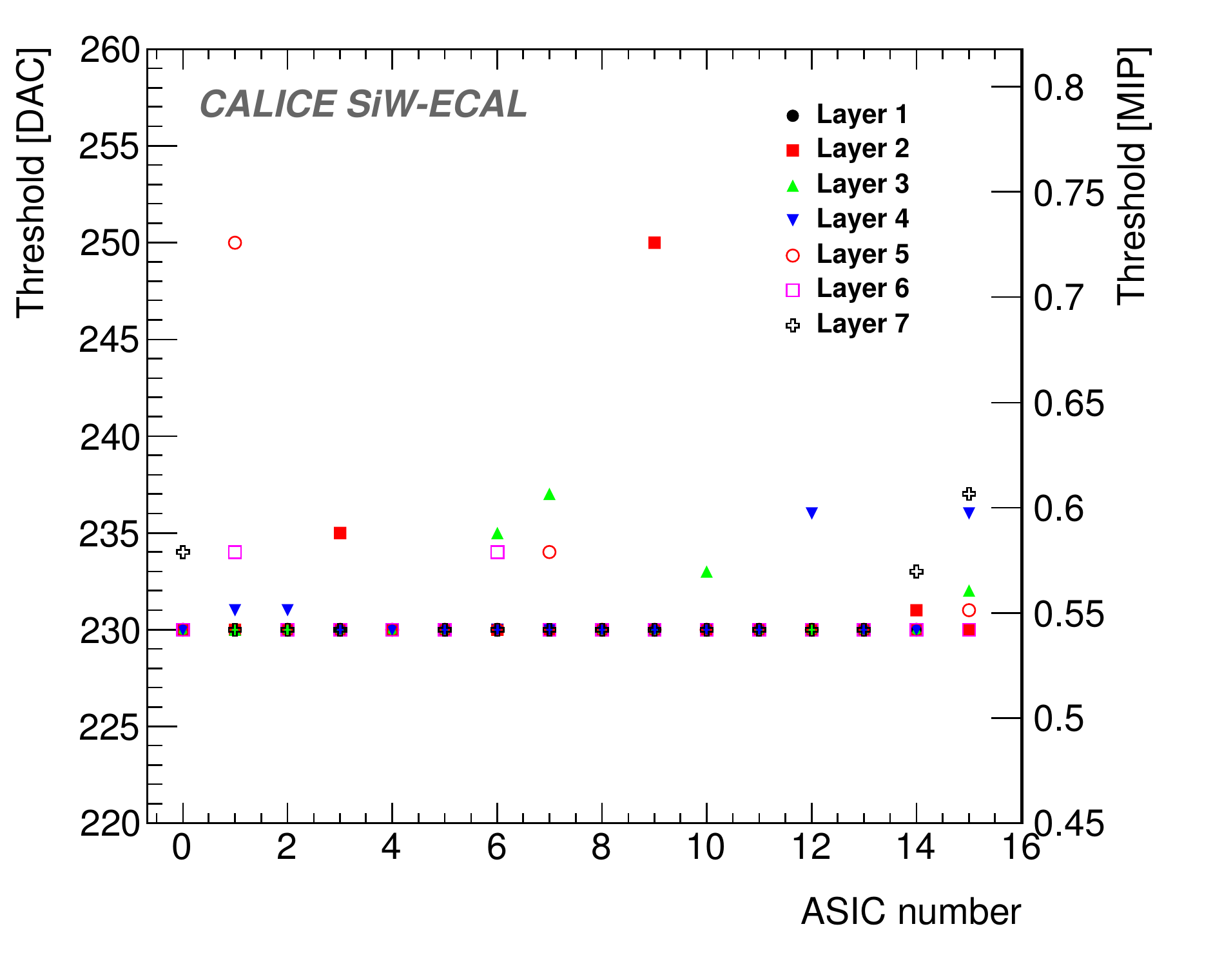}
  \caption{Summary of the trigger threshold settings in internal DAC units and in MIP units.}
\label{trigger_thresholds}
\end{figure}

From these two results we extract the value of
\begin{equation}
  S/N(trigger)=12.9\pm3.4
\end{equation}
for the trigger decision. The central value is calculated from Figure \ref{SKIROC2} by
the comparison of the 1 and 2 MIP curves and using the width of the 1 MIP curve in the denominator.
 The estimated
uncertainty has two components: the difference of width between the 1 and 2 MIP curves of injected signals and the differences (width and middle point) between the 1 MIP curves for injected and
cosmic ray signals. With this value of the 
S/N we are able to operate the internal trigger for small signals of the size of $\sim0.5$ MIP.
This is seen in \ref{trigger_thresholds} where the chosen thresholds of every ASIC being tested
in beam are shown. However dedicated studies in beam test are needed in order to
reduce the uncertainty of this measurement.

\subsection{Prospects}
\label{sec:comm_prospects}

The commissioning procedure described above relies on very conservative decisions
due to the presence of unknown noise sources during largest of the commissioning phase.
These sources are
now well know and therefore a new noise commissioning procedure has been studied.
It will consist on an iterative algorithm that first
will identify and mask the channels
in which the number of triggers per channel will be compared with the number of expected triggers
assuming only cosmic rays as signal. This will allow us to have a
definition of the noise levels for each channel independently
instead of relatively to the total number of triggers recorded by
the ASIC. Finally, once the noisy channels
are identified, the threshold  are further optimized
with a last run for the identification of the residual noisy channels.

Using this new procedure we manage to reduce the number of masked channels by a factor of two
without any loss of performance, at least in the laboratory and using 3 of the 7 readout layers.
This new procedure will also be applied in the next beam test.
Also, in order to
optimize the commissioning of the detector,
we propose a new set of measurements in the next beam test such as
a threshold scan for the determination of the S/N in the trigger line. The later
can be done by the comparison of threshold curves taken with incident MIP-acting particles
and MIP-acting particles traversing the detector tilted by 45 degrees with respect to the beam direction.

\section{Performance in a beam test with positrons at DESY}
\label{sec:beamtest}

The beam line at DESY provides continuous positron beams in the energy range of 1 to 6 GeV with
rates from a few hundreds of Hz to a few kHz with a maximum of $\sim 3$ kHz for 2-3 GeV. 
In addition, DESY gives access to a bore 1 T solenoid, the PCMag.

The physics program of the beam test can be summarized in the following points:

\begin{enumerate}
\item Calibration without tungsten absorber using 3 GeV positrons acting MIPs directed to 81 position equally distributed over the modules.
\item Test in magnetic field up to 1 T using the PCMag. For this test a special PVC structure was
  designed and produced to support one single readout layer.	
  The purpose of such test was twofold: first to prove that the DAQ, all electronic devices and the 
  mechanical consistency of the readout layer itself are able
  to handle strong magnetic fields; 
  second to check the quality of the data and the performance of the detector during the data taking when running
  in a magnetic field. 
  \item Response to electrons of different energies with fully equipped detector, i.e. sensitive parts {\it and} W absorber, with three different repartitions of the absorber material:
  \begin{itemize}
  \item W-configuration 1: $0.6,1.2,1.8,2.4,3.6,4.8$ and $6.6~X_{0}$
  \item W-configuration 2: $1.2,1.8,2.4,3.6,4.8,6.6$ and $8.4~X_{0}$
  \item W-configuration 3: $1.8,2.4,3.6,4.8,6.6,8.4$ and $10.2~X_{0}$
  \end{itemize}
\end{enumerate}

First reports on this beam test can be find in
Refs. \cite{Irles:2018uum,Irles:2018hcd}. These results have extended and
are discussed in the following sections. In Section \ref{sec:calib} we discuss in detail
the results of the pedestal, noise and MIP calibration.
We show also results on the pedestal and noise stability when running inside
a magnetic field in Section \ref{sec:magnetic} and in electromagnetic shower
events in Section \ref{sec:showers}. The study of the calibration of the prototype
in electromagnetic shower events is due to a future publication.

\subsection{Noise study and MIP calibration}
\label{sec:calib}

\begin{figure}[!ht]
  \centering
  \begin{tabular}{l}
    \includegraphics[width=4in]{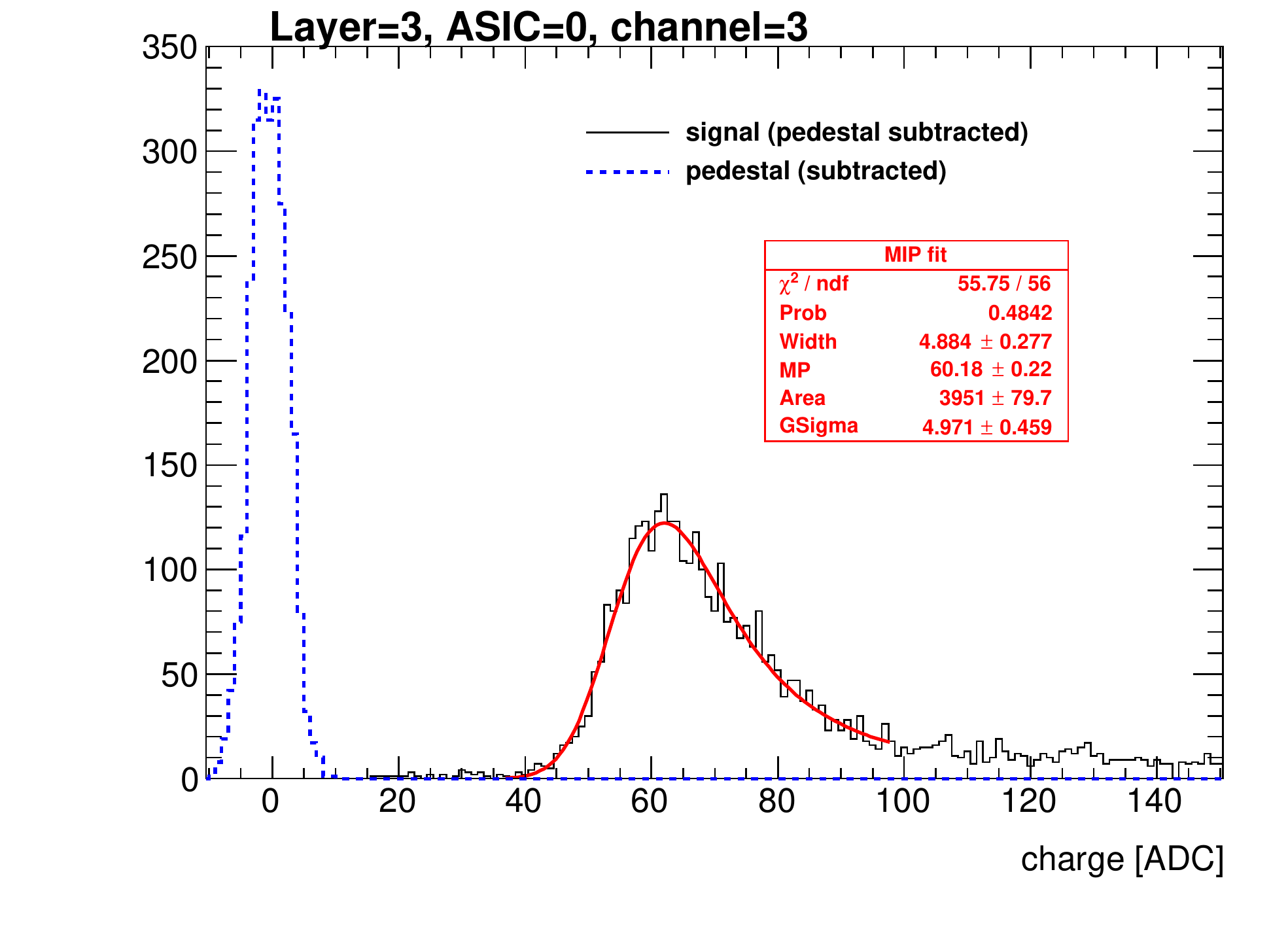}
  \end{tabular}
  \caption{Pedestal (blue dashed line) and signal (black continuous line) distribution for one channel in the third layer.}
\label{signal_pedestal}
\end{figure}

In Figure \ref{signal_pedestal} we show the signal and pedestal distribution of a single channel after
subtracting the pedestal mean position. The results of the MIP calibration fit are shown in red.
The signal distribution is integrated over all SCAs.
For cosmetic reasons the pedestal distribution is shown only for the first SCA.


\begin{figure}[!t]
  \centering
  \begin{tabular}{ll}
    \includegraphics[width=2.8in]{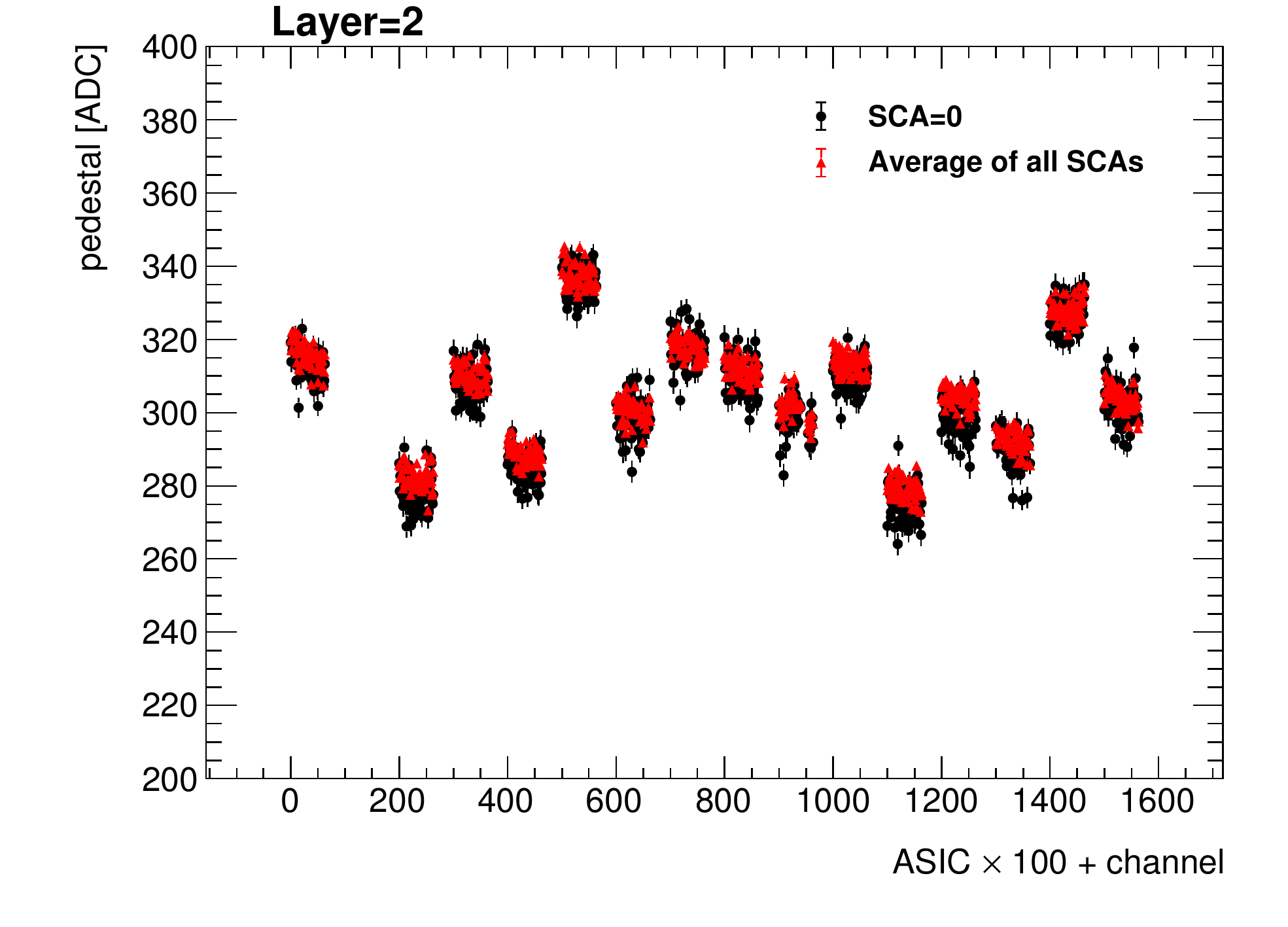} & \includegraphics[width=2.8in]{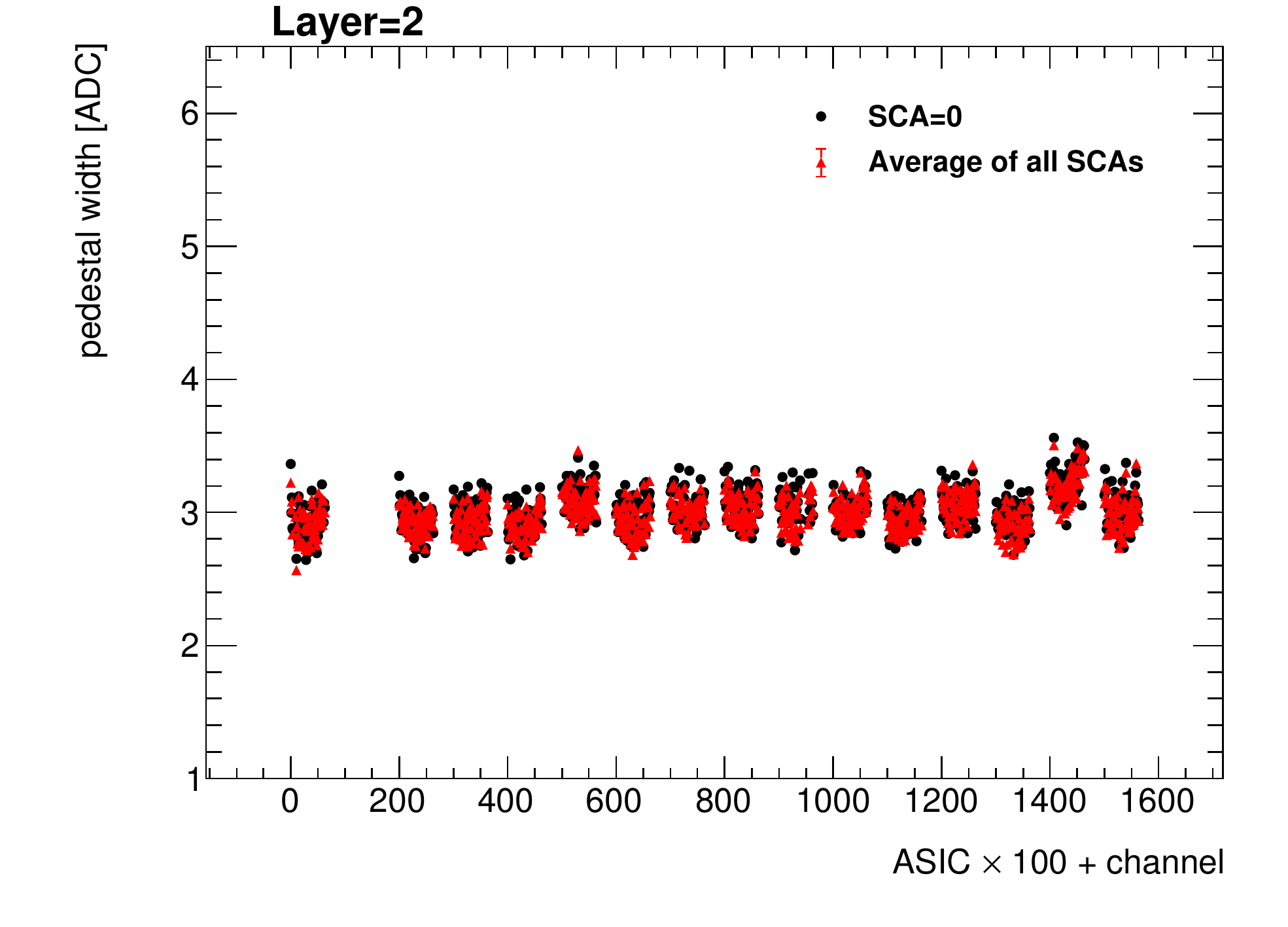} 
  \end{tabular}
  \caption{Upper plots: pedestal mean position (left plot) and width (right plot) for all channels in one layer. Lower plots: same plots but after subtracting the corresponding averaged values calculated for each ASIC and SCA.}
\label{pedestal_layer}
\end{figure}

\begin{figure}[!t]
  \centering
  \begin{tabular}{ll}
    \includegraphics[width=2.8in]{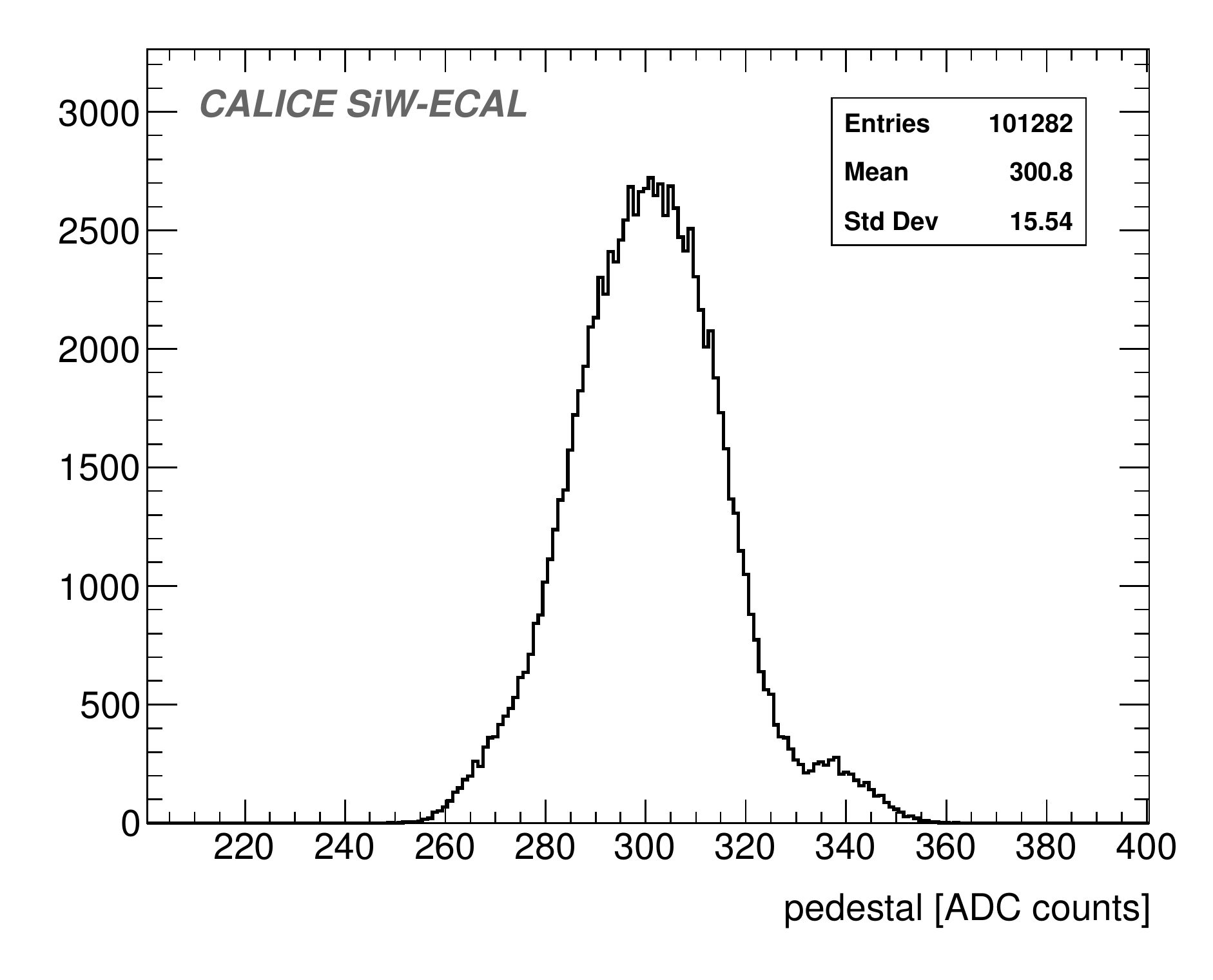} & \includegraphics[width=2.8in]{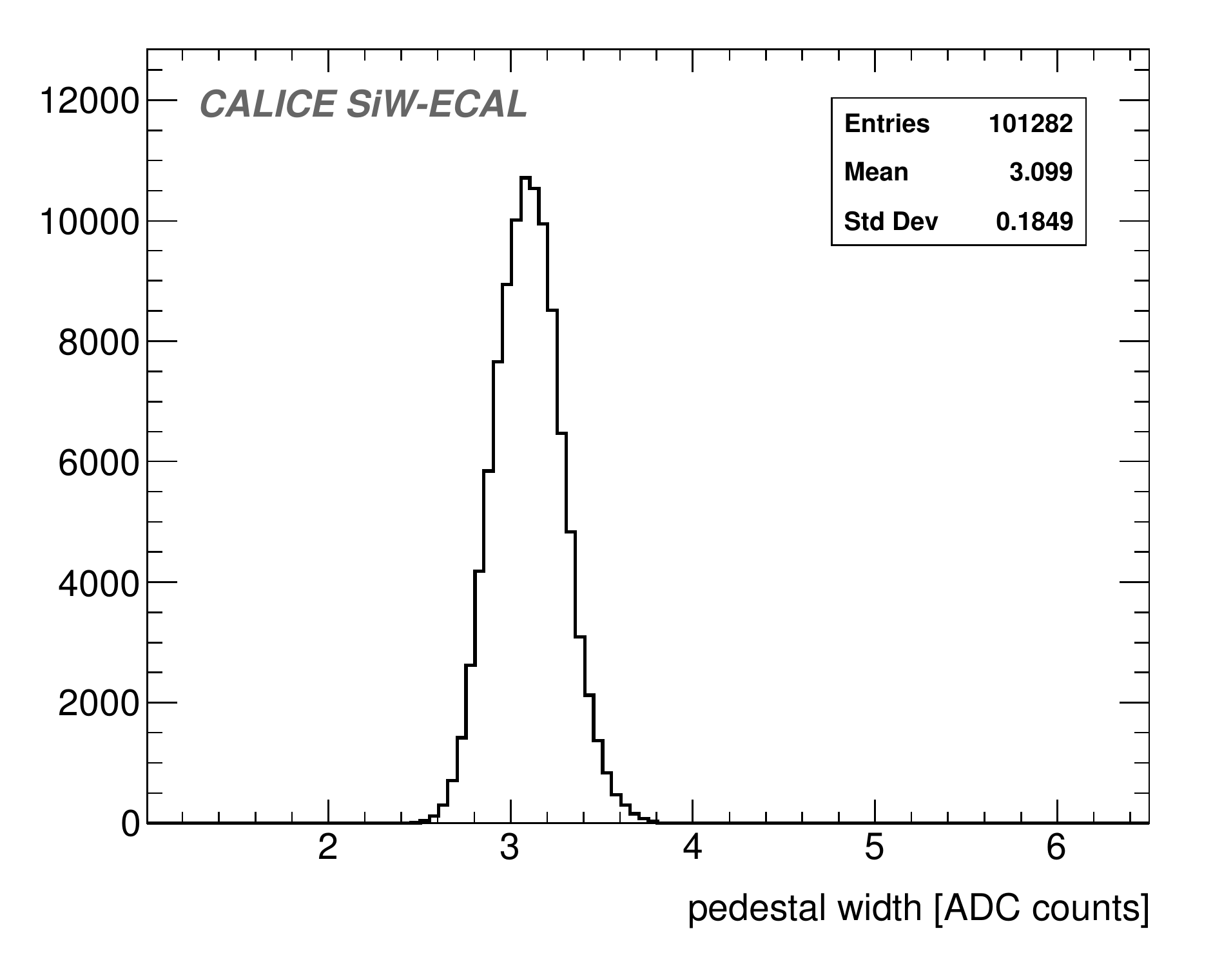} 
  \end{tabular}
  \caption{Pedestal mean position (left) and width (right) for all channels and all SCAs in the setup.}
\label{pedestal_all}
\end{figure}

\begin{figure}[!t]
  \centering
  \begin{tabular}{ll}
      \includegraphics[width=2.8in]{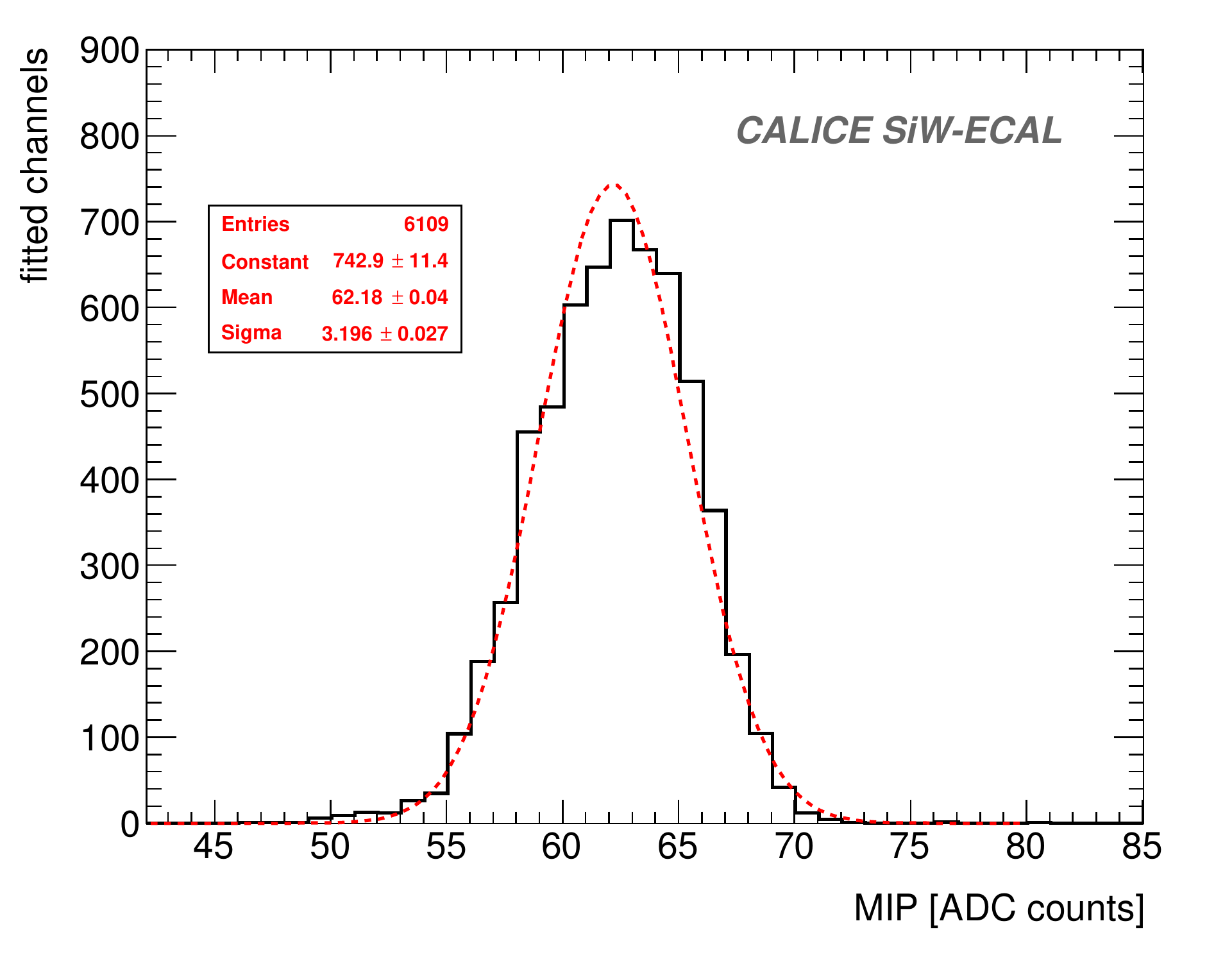} & \includegraphics[width=2.8in]{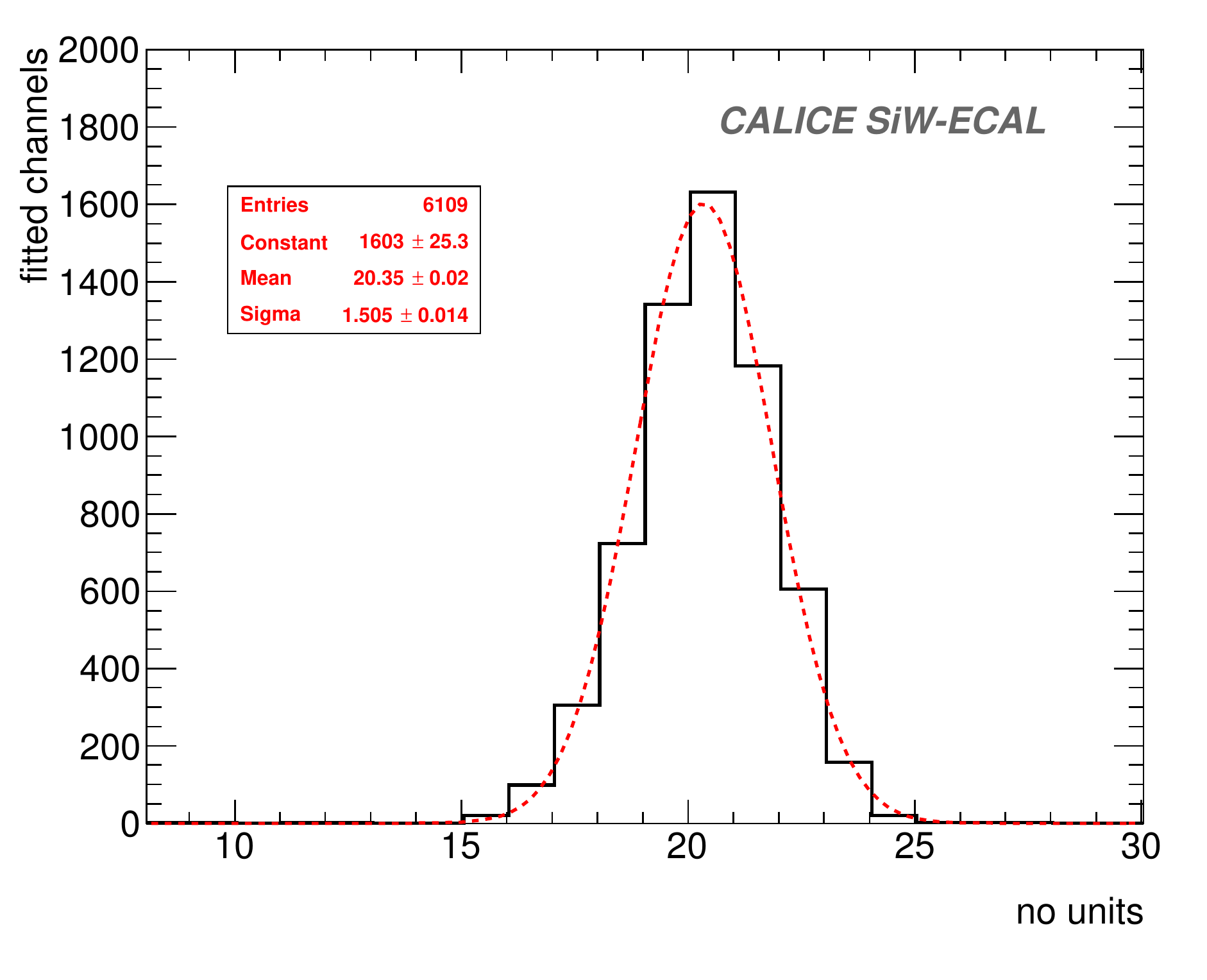}  
  \end{tabular}
\caption{Result of the MIP position calculation and signal over noise for the charge measurement on triggered channels for all calibrated channels.}
\label{mipandSN}
\end{figure}

The pedestal is calculated as the mean position of
the distribution of the ADC values for all channels without trigger. The noise is
associated to the width of the distribution.
The pedestal correction is done layer-, chip-, channel- and SCA-wise due to the large spread of values between pedestals, as observed in 
Figure \ref{pedestal_layer} (left plot) and Figure \ref{pedestal_all} (also left plot).
For the noise, the dispersion is much smaller ($\sim 5 \%$). This is shown in the right plots of Figures \ref{pedestal_layer} and \ref{pedestal_all}.
From now on, the pedestal correction is applied to all the results presented.
The resulting spectra are fit by
a Landau function convoluted with a Gaussian.
The most-probable-value of the convoluted function is taken as the MIP value, allowing thus for a direct
conversion from ADC units to energy in MIP units.
The fit succeeded in 98\% of the cases and the spread of the resulting MPV is 5\%.
The remaining channels will be discarded. Results are summarized in figure \ref{mipandSN}, leftmost plot.

\begin{figure}[!t]
  \centering 
    \begin{tabular}{ll}
      \includegraphics[width=4in]{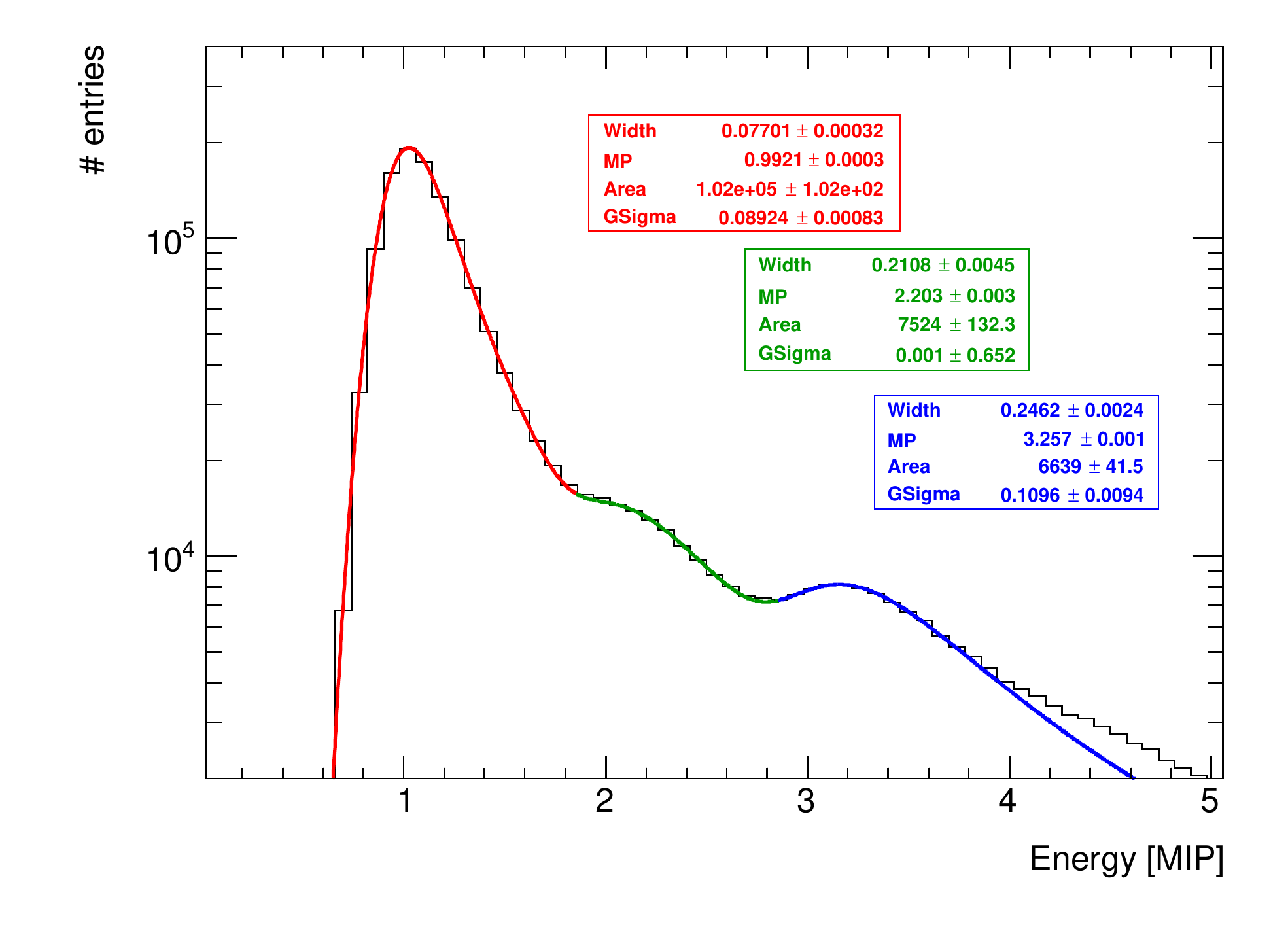} 
    \end{tabular}
    \caption{Energy distribution for all calibrated channels when selecting incident tracks of 3 GeV positron acting as MIPs.}
\label{mip3peaks}
\end{figure}

The Figure \ref{mip3peaks} shows the response of all channels integrated over the calibration run.
This plot is obtained after further refinement of the sample by selecting incident tracks.
The maximum peaks at 1 MIP as expected after a good calibration.
In addition to this, a second and a third peaks
are visible as shoulders. These shoulders are associated to 
events involving multiple 
particles crossing the detector.

\begin{figure}[!t]
  \centering 
  \includegraphics[width=4in]{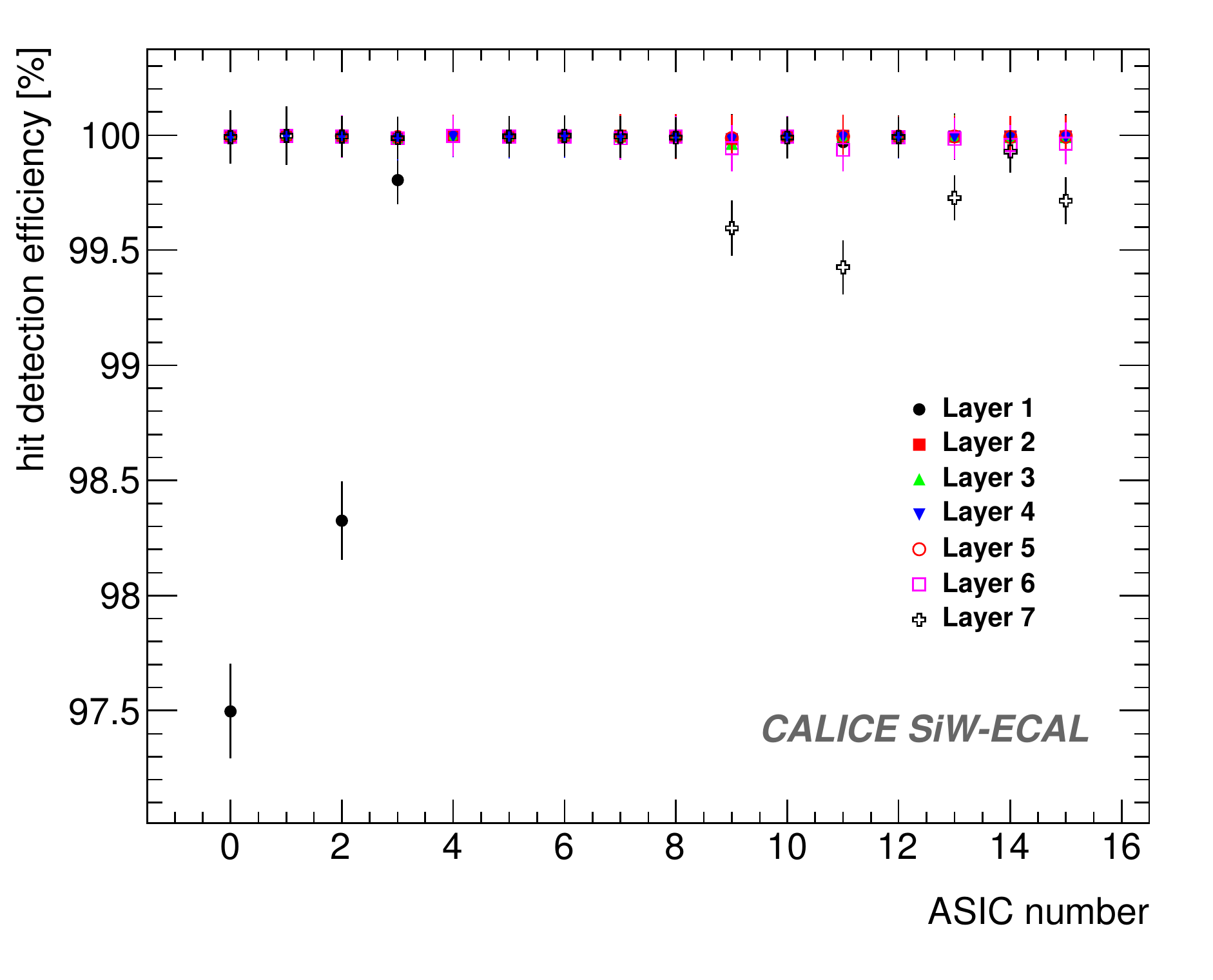}
  \caption{MIP detection efficiency for all layers and ASICs in high purity samples of tracks of MIP-like acting particles.}
\label{efficiency}
\end{figure}

To evaluate 
the single hit detection efficiency we define a high purity sample of
events by selecting
tracks with at least 4 layers with a hit in exactly the same channel. Afterwards we 
check which layers have or not a hit in the same or in the closest neighboring channels with energy larger or equal than 0.3 MIP.
We repeat this procedure for all channels.
The results are shown in Figure \ref{efficiency}. Except few exceptions, the efficiency is 
compatible with $100\%$.
The low efficiencies in the first layer are related to the presence of
noisy channels not spotted during the commissioning. These channels
may saturate de acquisition in their ASICs. In the last layer we also observe a few small deviations
which are associated to the outliers channels, hinting for a small misalignment of the last layer.

\subsubsection{S/N for the charge measurement of triggered channels}
\label{sec:sn}

The S/N for the charge measurement of triggered channels is defined 
as the ratio between the most-probable-value of
the Landau-gauss function fit to the data (pedestal subtracted) and the noise (the pedestal width). This quantity 
has been calculated for all channels and all layers. 
Results are summarized in Figure \ref{mipandSN}, rightmost plot.
The large size of this value allows us to filter out small spurious triggers 
with high efficiency.

\subsection{Pedestal and noise stability in a magnetic field}
\label{sec:magnetic}

The data taking inside the magnetic field has been divided in three runs:
\begin{enumerate}
\item a with a magnetic field of 1 T;
\item a run with 0.5 T;
\item and a final run with the magnet off.
\end{enumerate}

\begin{figure}[!t]
  \centering
  \begin{tabular}{ll}
    \includegraphics[width=2.8in]{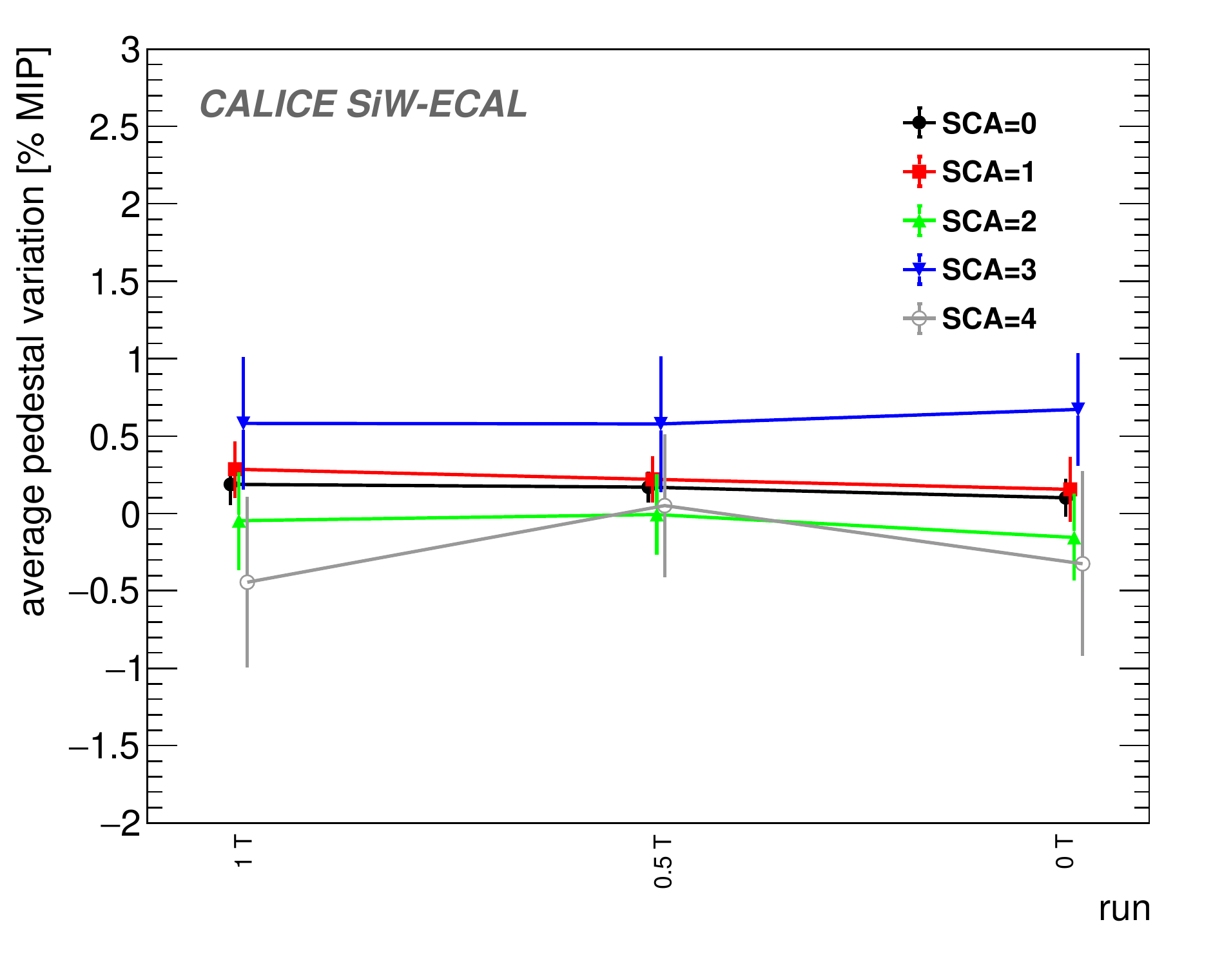} & \includegraphics[width=2.8in]{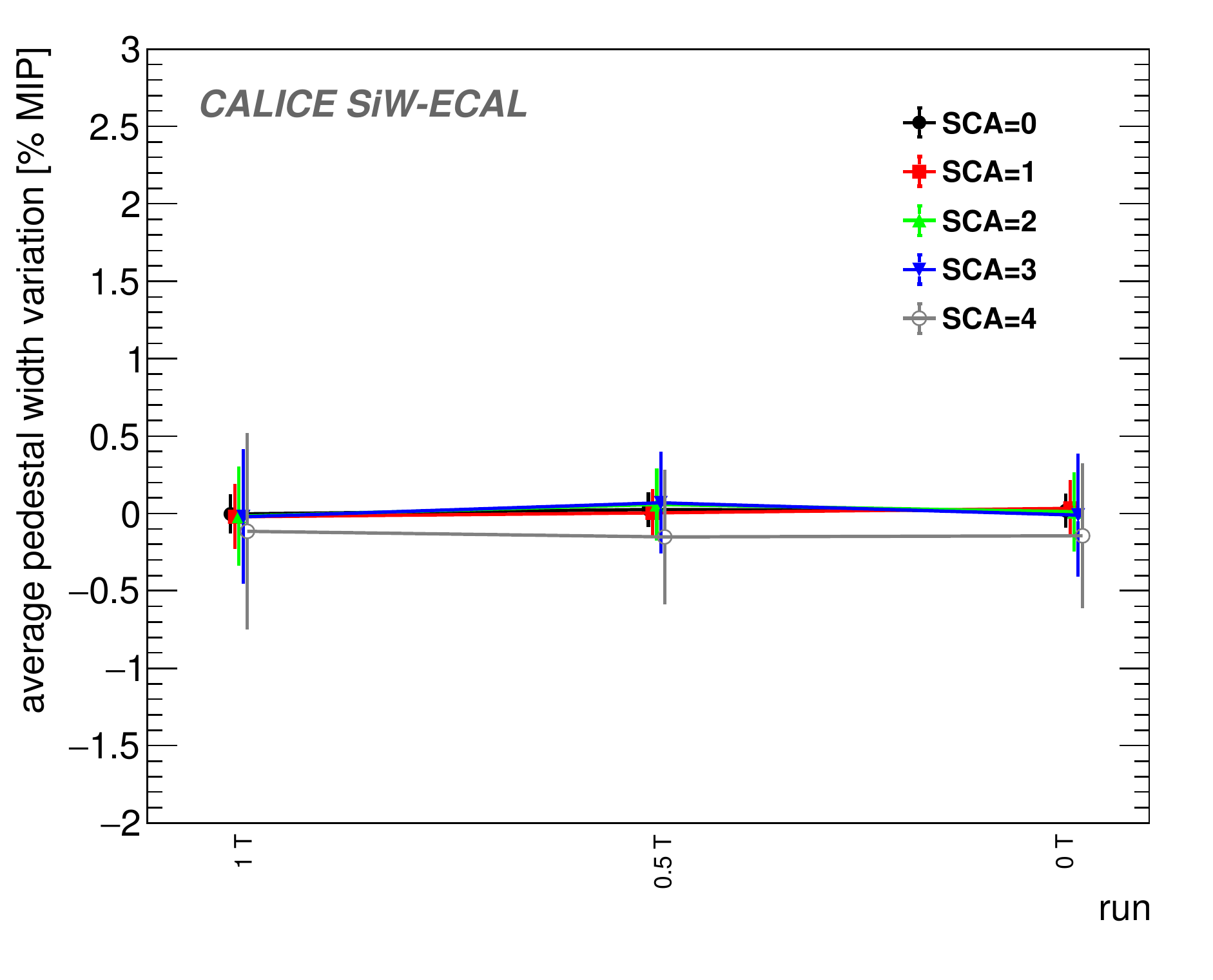}
  \end{tabular}
  \caption{Average deviation of the pedestal mean position (left) and width (right) for all channels in the ASIC 12.}
\label{pedestal_magnetic}
\end{figure}

The beam, 3 GeV positrons, was directed in the area of the PCB readout by the ASIC number 12.
The pedestal positions and noise levels of the channels of the ASIC 12 when the
readout layer is inside of the PCMag are compared with the results from the calibration run described in the previous section.
This is shown in Figure \ref{pedestal_magnetic}.
We see that the agreement is perfectly good within the statistical uncertainties.
Due to the lower rates in this beam area, the
analysis is only done up to few SCAs.

\subsection{Pedestal stability in electromagnetic shower events}
\label{sec:showers}

In this section we discuss the pedestal stability in events with large amount of charge collected by the ASICs, as are the 
electromagnetic shower events. All the results shown in this section correspond to data taken during the tungsten program, 
using the W-configuration number 2 when shooting the beam in the area registered by the ASIC 12 (and partially in the 13). 
For simplicity, only information recorded by ASIC 12 will be shown. 
In order to select a high purity of
electromagnetic shower like the events, 
we used a simple criteria: select only events with at least 6 of the layers with at least a hit with E > 0.5 MIP.

\begin{figure}[!ht]
  \centering 
    \begin{tabular}{ll}
      \includegraphics[width=2.8in]{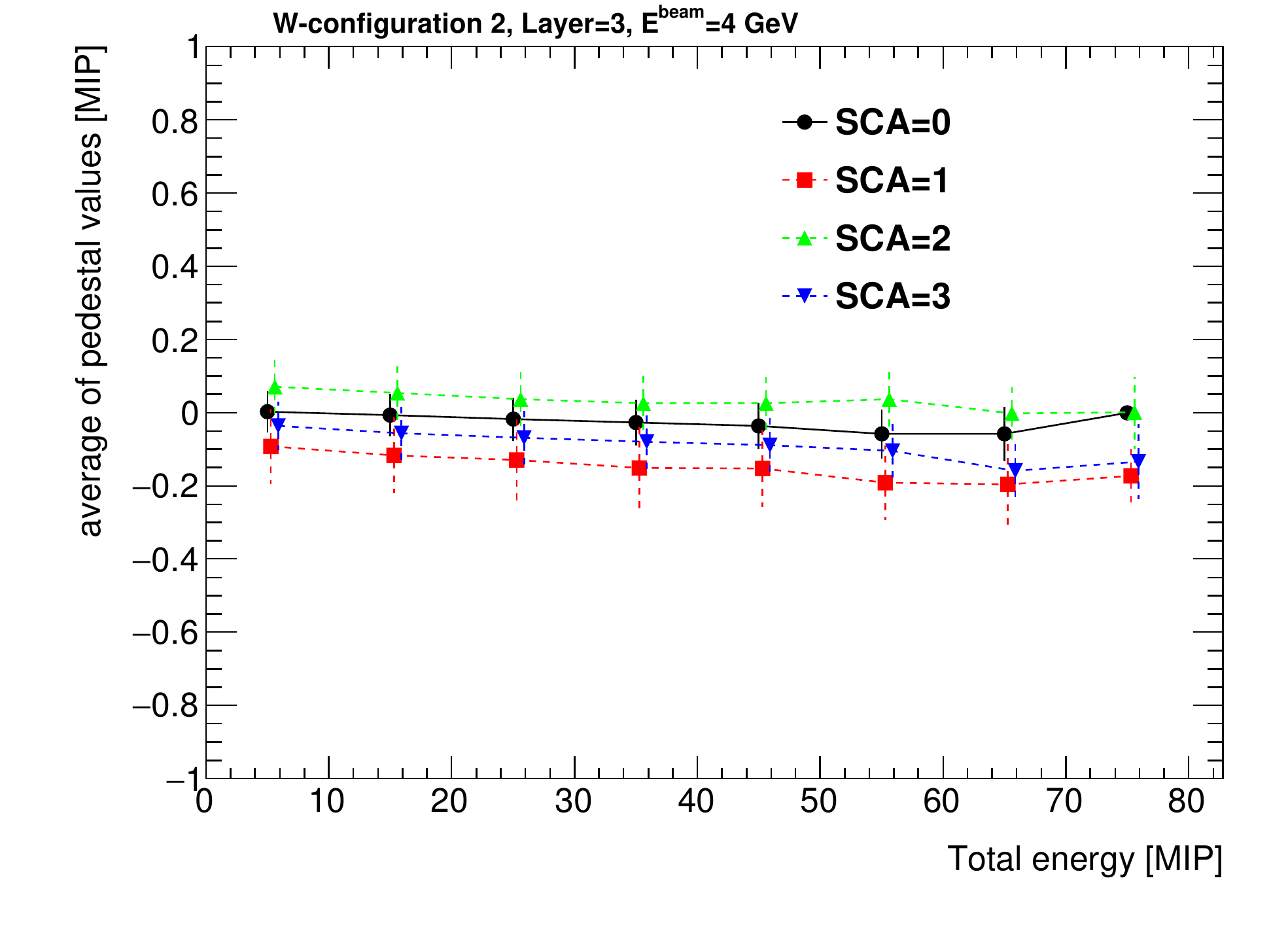} & \includegraphics[width=2.8in]{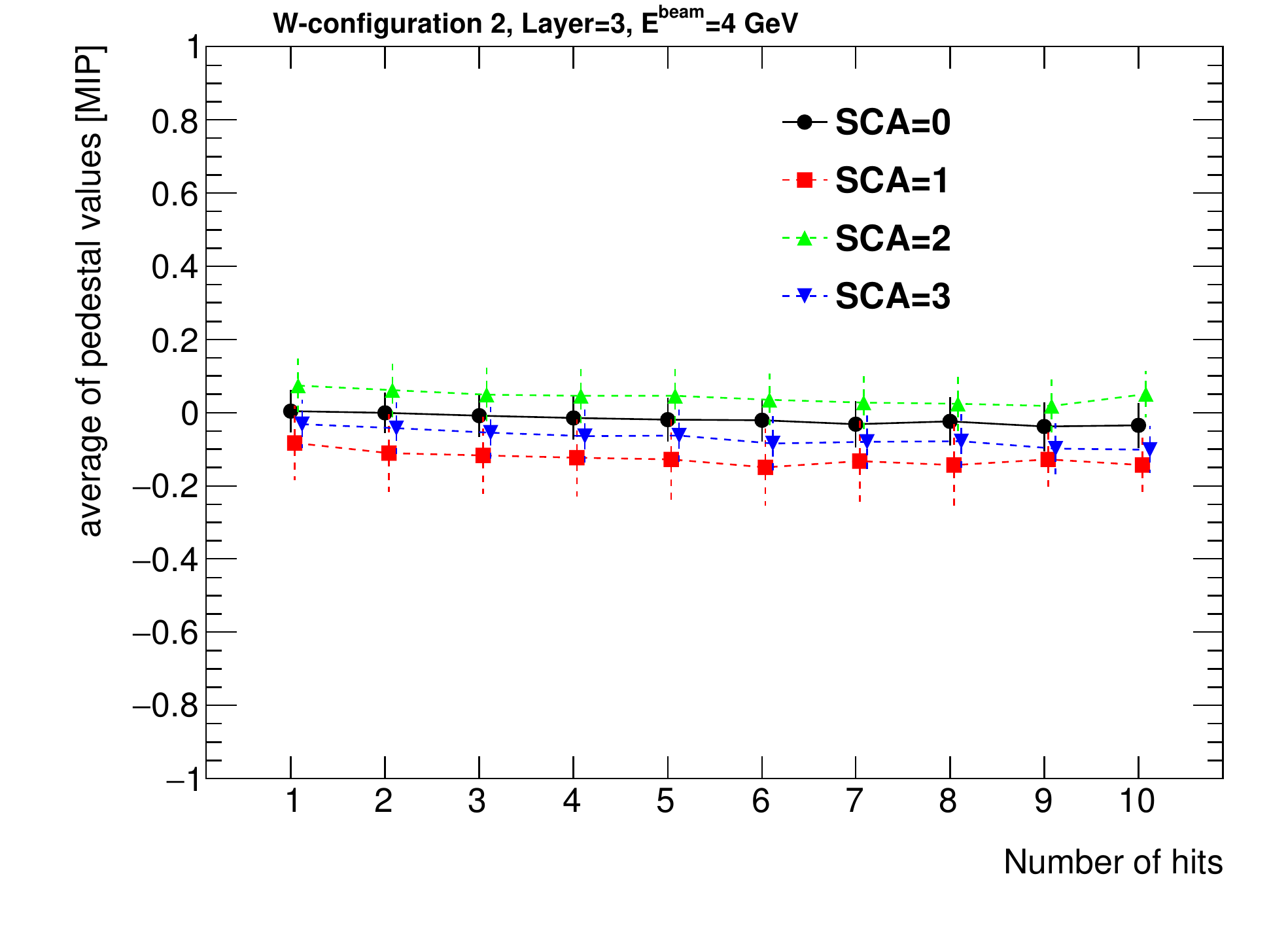} \\
    \end{tabular}
    \caption{Left: mean position of the projection of the pedestal distribution of all channels
      calculated when different energies are collected in the ASIC (in bins of 10 MIPs).
      Right: same but as a function of the number of hits. In both cases, the results are shown for few SCA.
      The points for the curves with SCA different than zero are slightly shifted in the x-axis to optimize the visualization.}
\label{pedestal_shower_1}
\end{figure}

\begin{figure}[!t]
  \centering 
    \begin{tabular}{ll}
      \includegraphics[width=2.8in]{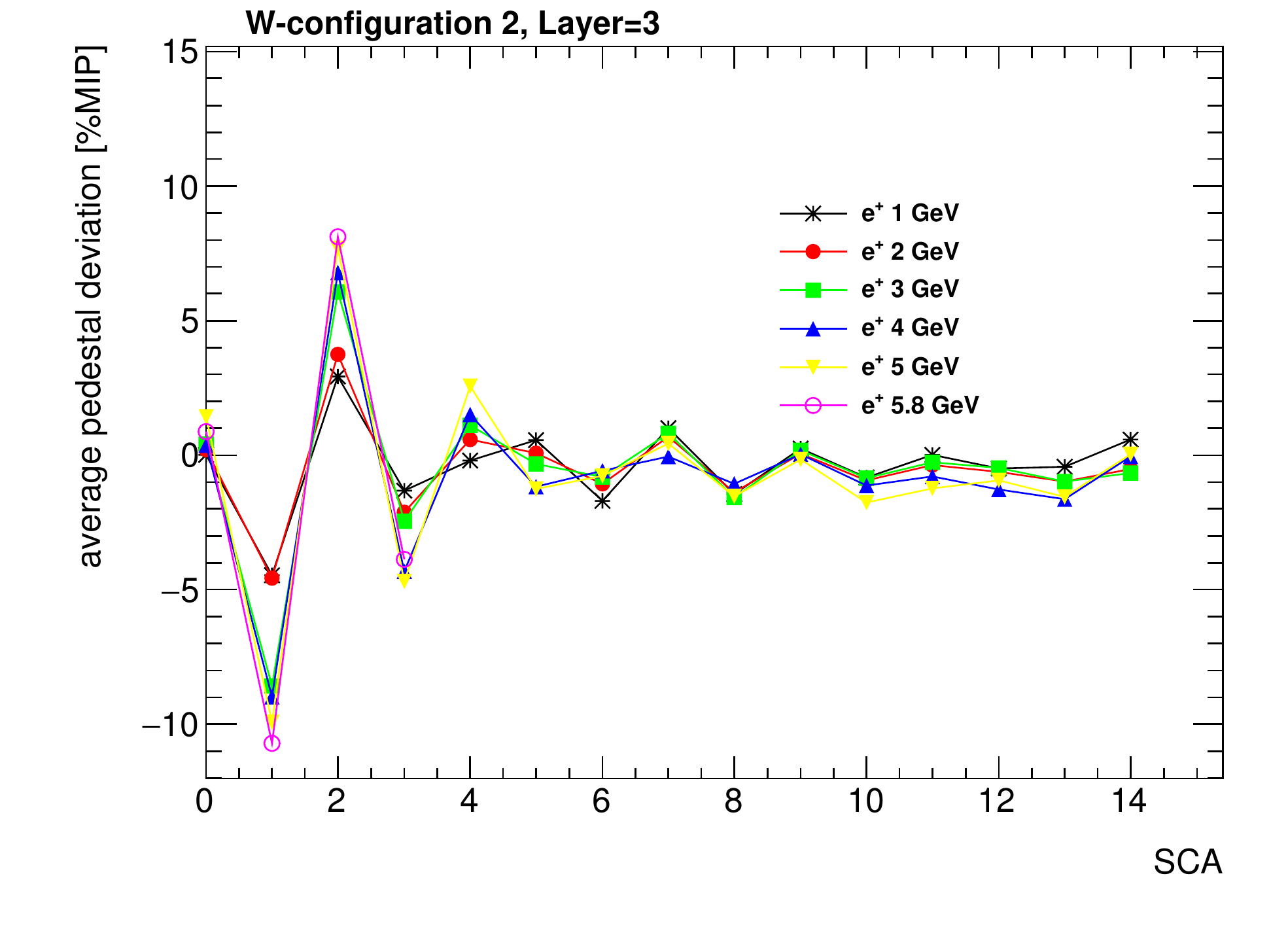} & \includegraphics[width=2.8in]{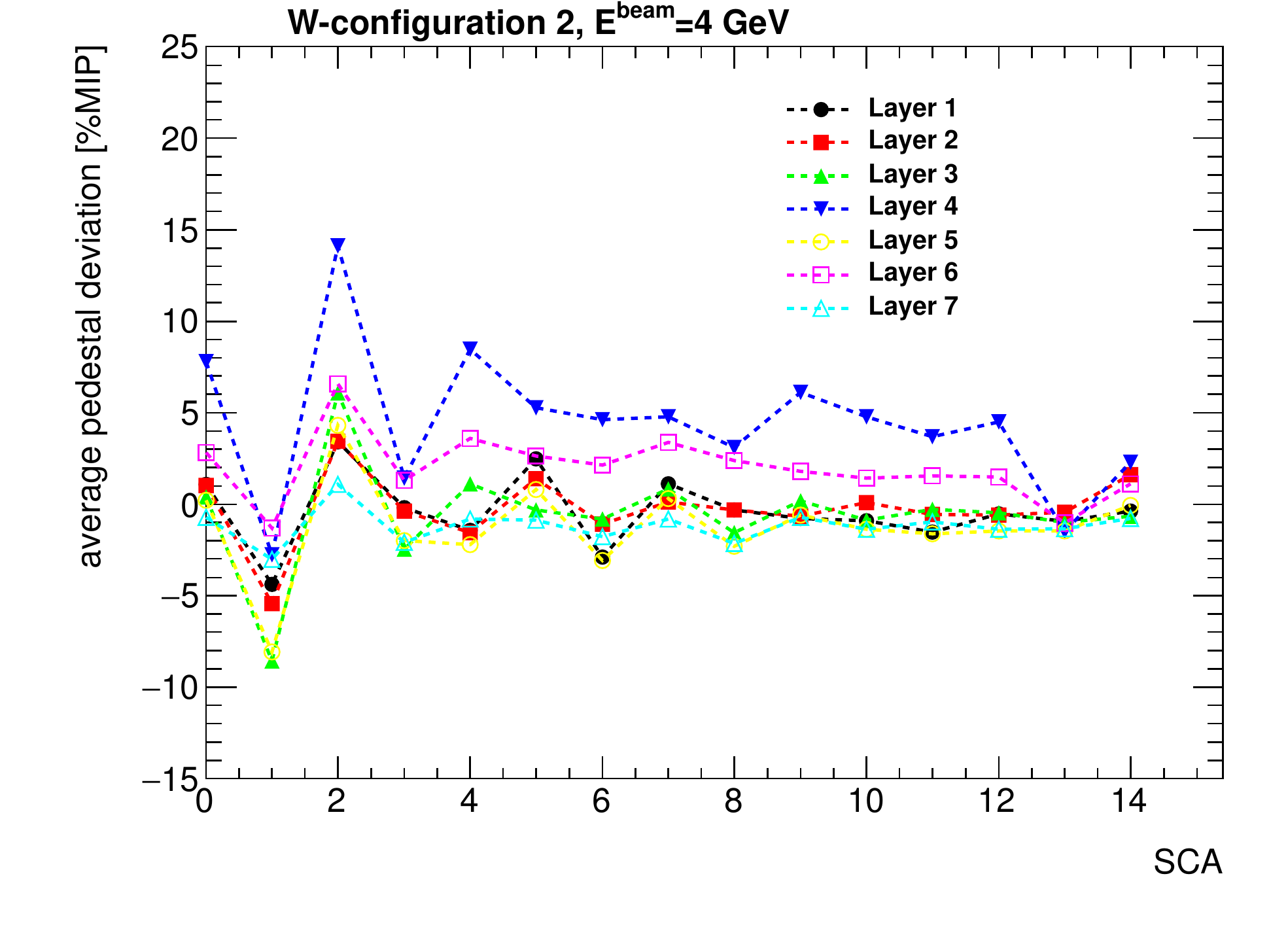} \\
    \end{tabular}
    \caption{Average value over all channels in ASIC 12 of the pedestals position for each SCA in electromagnetic shower events.}
\label{pedestal_shower_2}
\end{figure}

Two main observations have been extracted from the recalculation of the pedestals and its comparison
with the values obtained previously during the calibration runs. The first observation
consists in a relatively small 
drift of the pedestal values
towards lower values when the collected energy is high {\it i.e.} when the number of triggered channels is large.
This is shown in Figure \ref{pedestal_shower_1} for several SCAs.
A small dependence, in all SCAs, of the pedestal position on the amount of charge collected by the
ASIC is observed.
This feature is known and it is due to the architecture of the SKIROC2 ASICs 
where high inrush of currents can slightly shift the baseline of the analogue power supply. 
The second observation extracted from this analysis can be also seen in Figure \ref{pedestal_shower_1} but
more clearly in Figure \ref{pedestal_shower_2}: in addition
to the small drift of the pedestal value an SCA-alternate global shift
is observed. We see that the effect is enhanced when large amounts of charge
are deposited in the ASIC ({\it i.e.} at larger beam energies or for the layers in the maximum of the shower
profile). We also observed that this alternation is only SCA dependent and does not depends
on the time in which the deposit of energy occurs within the acquisition.
This is not yet fully understood although the fact that the effect is observed in
alternate SCAs hints that something is affecting to the digital part of the ASIC 
(where the SCAs enter in play).
Dedicated tests in the laboratory and in the beam are needed in order to clarify this issue.

\section{Summary}
\label{sec:summary}

The R\&D program of the highly granular SiW-ECAL detector is in an exciting phase. 
After the proof of principle of the imaging calorimetry concept using the physics prototype, the 
technological prototype is being constructed and tested. In this document we describe the commissioning and
beam test performance of a prototype built in with the first fully assembled
detector elements, in contrast with previous beam tests. In addition,
with the setup used in this beam test we reached levels of granularity
similar to the targets of the ILD detector for the ILC. This is also the first time
that results with the SiW-ECAL prototype continuously working in power pulsing mode are published.
Finally, we tested the performance of the detector
modules working for long periods inside magnetic fields.

The beam test has provided a lot of useful data to study 
the performance of the detector and to perform
a channel by channel calibration, showing a good homogeneity with a spread of the 5\% for all channels.
The S/N on the trigger has been evaluated to be $12.9\pm3.4$.
The S/N on the charge measurement of triggered cells is $20.4\pm1.5$. The hit detection efficiency
in tracks has been evaluated to be compatible with 100\%.

\section{Outlook}
\label{sec:outlook}

In parallel to the work described here, several R\&D efforts are being carried.
One of these efforts is directed to the design and test of new ASICs.
In fact, a new generation of the SKIROC, the 2a, has been delivered, tested in the dedicated testboards
and it has been integrated in new ASUs.
In addition, a new generation of the ASIC, SKIROC3, is foreseen for the final detector construction.
In contrast with SKIROC2/2a, the new ASIC will be fully optimized for ILC operation, {\it i.e.} full zero suppression, reduced power consumption etc.

Many efforts are also concentrated in the construction and test of long readout layers
made of several ASUs enchained since we know that the ILD ECAL will host long layers of up to $\sim$2.5m.
This device constitutes a technological challenge in both aspects, the mechanical
(very thin and long structure with fragile sensors in the bottom make complicated the assembly procedure and the handling...)
and the electrical (we need to ensure and control the transmission of signals and high currents along the full device).
For example, interconnections between ASUs and between ASU and interface card are one of
the most involved parts of the assembly
and require close collaboration between mechanical and electronic engineers.
A first long readout layer prototype
of $\sim8$ ASUs has been already tested in beam test also in DESY in 2018.

\begin{figure}[!t]
  \centering
    \includegraphics[width=2.8in]{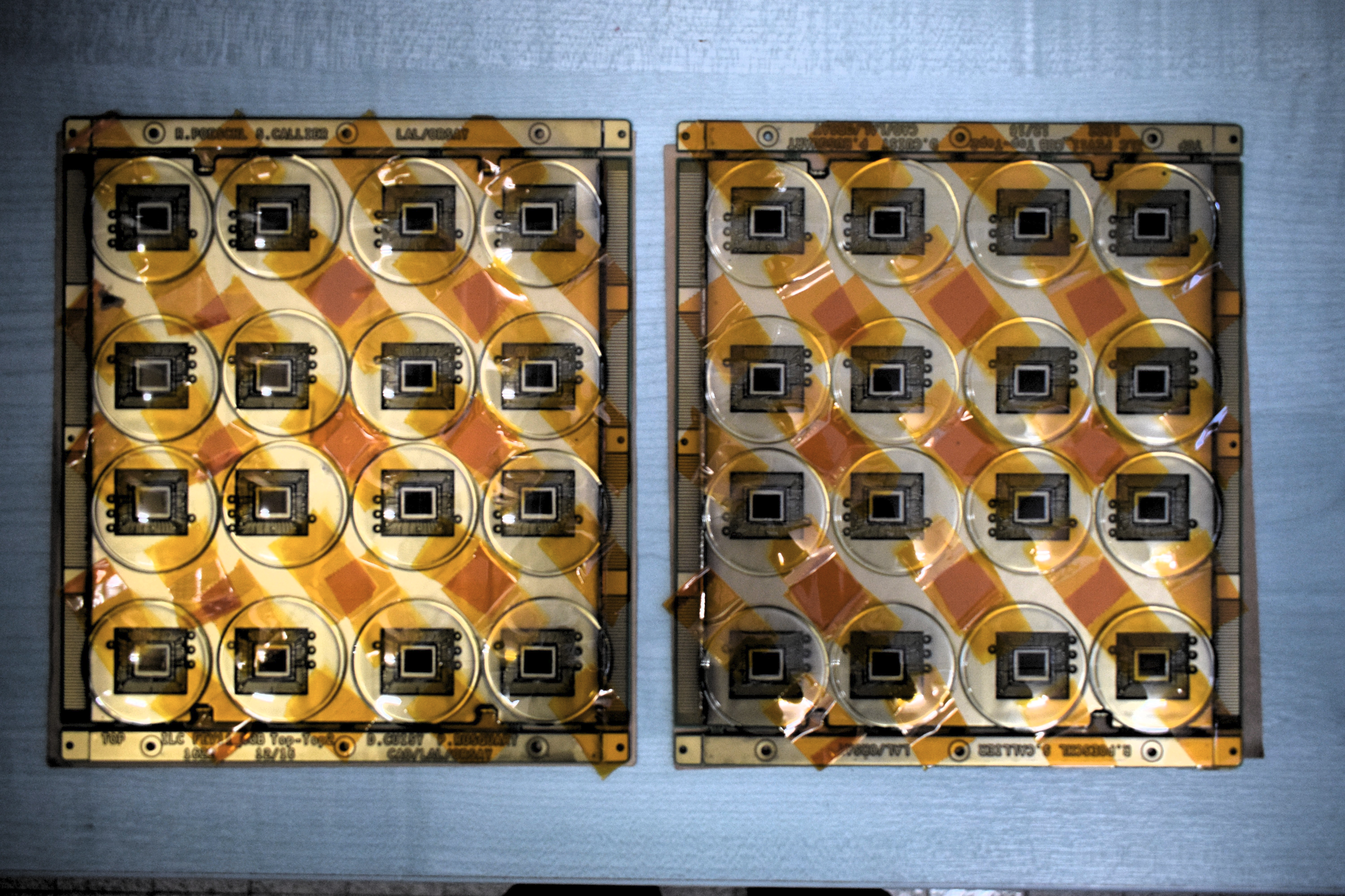} 
  \caption{Two FEV11\_COB boards with 16 SKIROC2a wire bonded. The ASICs are protected with watch glasses.}
\label{cob}
\end{figure}

In parallel, a different proposal for a thiner ASU
design is being investigated. This is motivated by the high density of channels
demanded by the Particle Flow algorithms. 
In this alternative PCB design the ASICs
are directly placed on board of the PCB in dedicated cavities.
The ASICS will be in semiconductor packaging and wire bonded to the PCB. This is the so-called COB (chip-on-board) version of the ASU.
A small sample of FEV11\_COBs (same connexion pattern with the interface card than FEV11)
with a total thickness of 1.2 mm (to be compared with the 2.7 mm of the LFBGA solution in the FEV11)
has been produced and tested in the laboratory
showing its readiness for tests with particle beams. A sample can be seen in Figure \ref{cob}.

Finally, intensive R\&D on the compactification of
the DAQ to meet the tight space requirements for the ILD is being done by the SiW-ECAL collaboration.

It is foreseen that all these developments, with the exception of the SKIROC3, will be tested with particle beams during 2018-2019.

\acknowledgments

This project has received funding from the European Union{\textquotesingle}s Horizon 2020 Research and Innovation program under Grant Agreement no. 654168.
This work was supported by the P2IO LabEx (ANR-10-LABX-0038), excellence project HIGHTEC,
in the framework {\textquotesingle}Investissements d{\textquotesingle}Avenir{\textquotesingle}
(ANR-11-IDEX-0003-01) managed by the French National Research Agency (ANR).
The research leading to these results has received funding from the People Programme (Marie
Curie Actions) of the European Union{\textquotesingle}s Seventh Framework Programme (FP7/2007-2013)
under REA grant agreement, PCOFUND-GA-2013-609102, through the PRESTIGE
programme coordinated by Campus France.
The measurements leading to these results have been performed at the Test Beam Facility at DESY Hamburg (Germany), a member of the Helmholtz Association (HGF).

\appendix
\section{Apendix: Filtering of fake triggers}
\label{sec:retriggers}

Several types of fake signals have been observed in the technollogical prototype since its construction and test. A detailed description of them
can be found in previous articles, as for example, in Ref. \cite{Amjad:2014tha}. All these fake signals are easily identified
and tagged during the data acquisition and removed afterwards from the analysis
not introducing any significance loss of performance as can be seen, for example,
in the hit detection efficiency plots (see Section \ref{sec:calib}).
In the following, we briefly describe the status of the monitoring, debugging and filtering
of such kind of events.

\subsubsection*{Empty triggers}

Empty trigger events are a well known feature of SKIROC2. The SKIROC2 uses
an OR64 signal to mark the the change to a new SCA when a signal over threshold is
detected. The empty triggers appear when during
the acquisition the rising edge of the slow clock falls during the OR64 signal
and therefore the change to a new SCA is validated twice.
This effect creates around 17\% of empty events which are easily filter and removed from the
analysis. The ratio of empty triggers in the new SKIROC2a has been reduced to the $\sim2-3\%$
by reducing the length of the OR64 signal.

\subsubsection*{Plane events and retriggers}

Another well know issue is the appearance of bunches of consecutive fake triggers, called retriggers,
that saturates the DAQ. 
Although the ultimate reason of the appearance of these events remains not clear, it
is suspected that they are related to distortions of the
power supply baselines. We know that the SKIROC2 and 2a preamplifiers are referenced to the analog power supply level,
therefore, any voltage dip can ve seen as signal by the preamplifiers.
Moreover the presence of a high inrush of current
due to many channels triggered at the same time can create these voltage dips
and also produce the so called plane events (most of the channels trigered at once).
In previous studies the ratio of retriggers and plane events
was reduced by improving the power supply stabilization capacitances. 

Studying the MIP calibration data of this beam test we have noticed that the 
concentration of the retriggers and plane events in
ASICs far from the beam spot is higher than in the ASICs that are reading
out the information of real hits. We have also observed that the concentration of these events
is higher in the nearby of channels that were masked as suspicious of suffering from routing issues.
The ratio these events have been estimated to be of $1-3\%$ in the ASICs where high frequency
interactions are produced ({\it i.e.} using 3 GeV positrons ate 2-3 kHz) and at higher rates even larger than $40\%$ in other 
ASICs far from the beam spot. 
Moreover, it has been noticed a correlation between the time that an ASIC was full and the time of the appearance of some 
retriggers in other areas of the PCB. 
This correlation corresponds to $\sim$1.6 $\mu$s which hints
of a distortion on the analogue power supply when
the signal that informs the DIF that one ASIC memory is full is transmitted through the PCB.



\bibliographystyle{JHEP}
\bibliography{../../references}

\end{document}